\documentclass[pra,aps,twocolumn,showpacs,amsmath,amssymb]{revtex4}

\usepackage{color}
\usepackage[dvips]{epsfig}
\usepackage{bbm}
\usepackage{subfigure}

\definecolor{orange}{rgb}{1,0.6,0}

\newcommand{\Hb}{\protect{{\bf H}}}
\newcommand{\Eb}{\protect{{\bf E}}}
\newcommand{\Pb}{\protect{{\bf P}}}
\newcommand{\Mb}{\protect{{\bf M}}}
\newcommand{\Bb}{\protect{{\bf B}}}

\newcommand{\rb}{\protect{{\bf r}}}
\newcommand{\nb}{\protect{\bf n}}
\newcommand{\db}{\protect{\bf d}}
\newcommand{\kb}{\protect{{\bf k}}}
\newcommand{\fom}{F\!o\!M}
\newcommand{\ep}{\protect{{\bf e}_+}}
\newcommand{\ex}{\protect{{\bf e}_x}}
\newcommand{\ey}{\protect{{\bf e}_y}}
\newcommand{\ez}{\protect{{\bf e}_z}}
\newcommand{\eins}{\protect{\mathbbm 1}}

\renewcommand{\Im}{\text{Im}}
\renewcommand{\Re}{\text{Re}}
\newcommand{\eq}[1]{Eq.~(\ref{#1})}
\newcommand{\bra}[1]{\ensuremath{\left<#1\right|}}
\newcommand{\ket}[1]{\ensuremath{\left|#1\right>}}

\newcommand{\ketbra}[2]{\ensuremath{\left|#1\right>\left<#2\right|}}

\begin{document}

\title{Low-loss negative refraction by
laser induced magneto-electric cross-coupling}
\author{J\"urgen K\"astel and Michael Fleischhauer}
\affiliation{Fachbereich Physik, Technische Universit\"{a}t
Kaiserslautern, D-67663 Kaiserslautern,
Germany}

\author{Susanne F. Yelin}
\affiliation{Department Of Physics, University of Connecticut,
Storrs, Connecticut 06269, USA}
\affiliation{ITAMP, Harvard-Smithsonian Center for Astrophysics,
Cambridge, Massachusetts 02138, USA}

\author{Ronald L. Walsworth}
\affiliation{Harvard-Smithsonian Center for Astrophysics and Department
of Physics, Harvard University, Cambridge, Massachusetts 02138, USA}

\begin{abstract}
We discuss the feasibility of negative refraction with reduced absorption in coherently driven
atomic media. Coherent coupling of an electric and a magnetic dipole transition by laser fields induces
magneto-electric cross-coupling and negative refraction at dipole densities which are considerably
smaller than necessary to achieve a negative permeability. At the same time 
the absorption gets minimized due to destructive quantum interference and the ratio
of negative refraction index to absorption becomes orders of magnitude larger than 
in systems without coherent cross-coupling. The proposed scheme allows for a fine-tuning 
of the refractive index. We derive a generalized expression for the impedance of
a medium with magneto-electric cross coupling and show that impedance matching to vacuum
can easily be achieved. Finally we discuss the tensorial properties of the medium response
and derive expressions for the dependence of the refractive index on the propagation direction.
\end{abstract}

\date{\today}

\maketitle

\section{Introduction}
\label{ch:introduction}

Negative refraction of light, first predicted to occur in materials with simultaneous negative
permittivity and permeability in the late 60's \cite{Veselago68}, has become one of the most active
fields of research in photonics in the last decade. Since the theoretical proposal for its realization
in meta-materials \cite{Pendry99,Smith00} and its first experimental demonstration \cite{Shelby01} 
for RF radiation, substantial technological progress has been made towards negative refraction
for shorter and shorter wavelengths \cite{Shalaev07,Soukoulis07}.
This includes various approaches based on split-ring resonator meta-materials
\cite{Shelby01,Yen04,Linden04,Enkrich05}, photonic crystals \cite{Parimi04,Berrier04,Lu05}
as well as more unconventional designs like double rod
\cite{Shalaev05,Klar06,Yuan07} or fishnet structures \cite{Zhang05,Dolling07}. 

Despite the wide variety of implementations a major challenge is the large loss rate
of these materials \cite{Shalaev07,Dolling07b}.
Especially for potential applications such as sub-diffraction limit imaging \cite{Pendry00}
or electromagnetic cloaking \cite{Leonhardt06,Pendry06,Schurig06}
the suppression of absorption proofs to be crucial \cite{Smith03,Merlin}.
The usually adopted figure of merit
\begin{equation}
\fom=-\frac{\Re[n]}{\Im[n]}
\label{eq:fom}
\end{equation}
reaches only values on the order of unity in all current meta-material implementations
\cite{Shalaev07} with a record value of $\fom=3$ \cite{DollingRekord}. This means that the
absorption length of these materials is only on the order of the wavelength.

Recently we have proposed a scheme in which coherent cross-coupling of an electric and a
magnetic dipole resonance with the same transition frequencies in an atomic system \cite{Scully-chirality} leads to negative refraction with strongly suppressed
absorption \cite{Kaestel07a} due to quantum interference effects similar to electromagnetically induced transparency
\cite{Harris-Physics-Today-1997,Fleischhauer05}. 
Furthermore, the value of the refractive
index can be fine-tuned by the strength of the coherent coupling.
In the present paper we provide a more detailed analysis
of this scheme. In particular we will discuss under what conditions a magneto-electric cross coupling 
can induce negative refraction in atomic media without requiring $\Re[\mu]<0$.
The model level scheme introduced in \cite{Kaestel07a} will be analyzed in detail. Explicit expressions for the
susceptibilities and cross-coupling coefficients will be derived and the limits of linear response theory
explored. The important issue of non-radiative broadenings and local field corrections to the response
will be discussed. Furthermore we will derive an explicit expression for the impedance of a medium
with magneto-electric cross coupling and show that it can be matched to vacuum via external laser fields
such that reflection losses at interfaces can be avoided. 
Finally we will give full account of the tensorial properties 
of the induced magneto-electric cross coupling and the resulting refractive index in the model system.
We will show that in the model system discussed it is in general only possible to obtain isotropic
negative refraction in 2D.


\section{Fundamental concepts}
\label{ch:concepts}

In the following we discuss the prospects of negative refraction in media
with magneto-electric cross-coupling. The electromagnetic constitutive relations 
between medium polarization $\Pb$ or magnetization $\Mb$ and the electromagnetic
fields $\Eb$ and $\Hb$ are usually expressed in terms on 
permittivity and permeability only. Important aspects of linear optical systems such as optical activity,
which describes the rotation of linear polarization in chiral media cannot be described in this
way however. The most general linear relations that also include these effects read
\begin{equation}
\begin{split}
\Pb= & \bar\chi_e\Eb+\frac{\bar\xi_{EH}}{4\pi}\Hb \\
\Mb= & \frac{\bar\xi_{HE}}{4\pi}\Eb+\bar\chi_m\Hb.
\end{split}
\label{eq:Pendrychiral}
\end{equation}
Eqs.(\ref{eq:Pendrychiral}) describe media with {\it magneto-electric cross coupling} which are also known
as bianisotropic media \cite{Kong1972}. Here the polarization $\Pb$ gets an additional
contribution induced by the magnetic field strength $\Hb$ and likewise the
magnetization is coupled to the electric field component $\Eb$.
$\bar\xi_{EH}$ and $\bar\xi_{HE}$ denote tensorial coupling coefficients
between the electric and magnetic degrees of freedom, while $\bar\varepsilon=1+4\pi\bar\chi_e$ and
$\bar\mu=1+4\pi\bar\chi_m$ are the complex-valued permittivity and permeability tensors.
As we use Gaussian units the coefficients $\bar\chi_e$, $\bar\chi_m$, $\bar\xi_{EH}$,
and $\bar\xi_{HE}$ are unitless.

The propagation properties of electromagnetic waves in such media are governed by the Helmholtz equation
\begin{equation}
\left[\bar{\varepsilon}+(\bar{\xi}_{EH}+\frac{c}{\omega}\kb\times)\bar{\mu}^{-1}(\frac{c}{\omega}
\kb\times-\bar{\xi}_{HE})\right]\Eb=0.
\label{eq:HelmholtzChiral}
\end{equation}
A general solution of this equation for the wave vector $\kb$ is very tedious \cite{ODell}
and a comprehensive discussion of the most general case almost impossible.
For the sake of simplicity we therefore assume the permittivity $\bar\varepsilon$ and
the permeability $\bar\mu$ to be isotropic $\bar\varepsilon=\varepsilon\eins$,
$\bar\mu=\mu\eins$. We furthermore restrict ourselves to a one-dimensional theory by choosing
the wave to propagate in the $z$-direction which leaves only the upper left $2\times2$-submatrices
of the tensors $\bar\xi_{EH}$ and $\bar\xi_{HE}$ relevant.
At this point we restrict the discussion to media which allow for conservation of the photonic angular momentum
at their interfaces. In particular, we assume the response matrices $\bar\xi_{EH}$ and $\bar\xi_{HE}$
to be diagonal in the basis $\{\ep,{\bf e}_-,\ez\}$
. Here ${\bf e}_\pm$ denote circular polarization basis
vectors ${\bf e}_\pm=(\ex\pm i\ey)/\sqrt{2}$. This leads, e.g., for $\bar\xi_{EH}$, to
\begin{equation}
\begin{split}
\bar\xi_{EH}= & \,\xi_{EH}^{+}\,\ep\otimes{\bf e}_+^\ast+\xi_{EH}^{-}{\bf e}_-\,\otimes{\bf e}_-^\ast+\xi_{EH}^{z}\,\ez\otimes\ez \\[2mm]
= & \left(\begin{array}{ccc}
(\xi_{EH}^{+}+\xi_{EH}^{-})/2 & -i(\xi_{EH}^{+}-\xi_{EH}^{-})/2 & 0 \\[2mm]
i(\xi_{EH}^{+}-\xi_{EH}^{-})/2 & (\xi_{EH}^{+}+\xi_{EH}^{-})/2 & 0 \\[2mm]
0 & 0 & \xi_{EH}^z
\end{array}\right).
\end{split}
\label{eq:TensorStructureXi}
\end{equation}
Note that the tensor (\ref{eq:TensorStructureXi}) includes also
biisotropic media first discussed by Pendry \cite{Pendry04} for 
negative refraction as a special case.
Such biisotropic media display a chiral behavior, i.e.,
isotropic refractive indices which are different for the two circular eigen-polarizations.
In section \ref{ch:2D} we will give an example which implements a polarization
independent but anisotropic index of refraction.

Using (\ref{eq:TensorStructureXi}) the propagation equation for 
a left circular polarized wave traveling in the $z$-direction can be expressed as
\begin{equation}
\varepsilon\mu-\left(\xi_{EH}^++i\frac{c}{\omega}k_z^+\right)\left(\xi_{HE}^+-i\frac{c}{\omega}k_z^+\right)=0
\end{equation}
which can be solved for $k_z^+$. As $k_z^+$ is related to the corresponding refractive index via $n^+=k_z^+c/\omega$
we find
\begin{equation}
n^+=\sqrt{\varepsilon\mu -\frac{\left(\xi_{EH}^{+}+\xi_{HE}^{+}\right)^2}{4}}+\frac{i}{2}\left(\xi_{EH}^{+}-\xi_{HE}^{+}\right).
\label{eq:refind}
\end{equation}
Note that for a vanishing cross-coupling ($\xi_{EH}=\xi_{HE}=0$) this
simplifies to the well known expression $n=\sqrt{\varepsilon\mu}$.

Equation (\ref{eq:refind}) has more degrees of freedom than the non-chiral version and allows for
negative refraction without requiring a negative permeability.
For example if we set $\xi_{EH}=-\xi_{HE}=i\xi$, with $\xi>0$ the refractive index (\ref{eq:refind})
reads
\begin{equation}
n=\sqrt{\varepsilon\mu}-\xi
\label{eq:refindsimple}
\end{equation}
Here and in the following we drop the superscript $^+$ for notational simplicity.

Magnetic dipole transitions in atomic systems are a relativistic effect
and thus magnetic dipoles are typically smaller than electric ones
by factor given by the fine structure constant $\alpha\approx 1/137$.
Since furthermore magnetic resonances are typically not radiatively broadened
the magnetic susceptibility $\chi_m$ per dipole is typically less than the electric 
susceptibility $\chi_e$ by a factor given by the fine structure constant squared $\alpha^2$. 
On the other hand, as we will
show later on, the cross-coupling coefficients $\xi_{EH}$ and $\xi_{HE}$ scale only
with one factor of $\alpha$. 
Thus \eq{eq:refindsimple} represents an improvement compared to
non-chiral approaches, since a negative index can be achieved at densities where
$\mu$ is still positive but $\xi>\sqrt{\varepsilon\mu}$.
Negative refraction with $\Re[\mu]>0$ comes with the requirements
to find large enough chiral coupling coefficients $\xi_{EH}$ and $\xi_{HE}$
and additionally to control their phases in order to get close enough to $\xi_{EH}=-\xi_{HE}=i\xi$ necessary
for {\it negative} refraction.

\begin{figure}[t]
     \begin{center}
       \includegraphics[width=5cm]{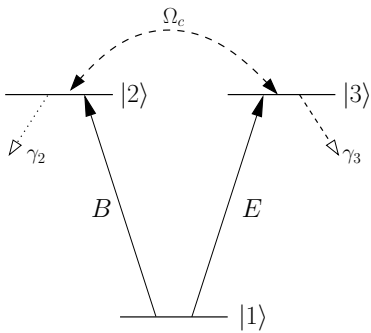}
       \caption{A simplistic three-level system which employs electro\-magnetically-induced cross-coupling. $E$, $B$ are the electric and magnetic components of the probe field, $\gamma_2$ and $\gamma_3$ the decay rates out of levels $|2\rangle$ and $|3\rangle$. $\Omega_c$ is an applied field which couples levels $|2\rangle$ and $|3\rangle$.}
       \label{fig:3level}
     \end{center}
\end{figure}

As in \cite{Kaestel07a} we will analyze these fundamental concepts in more detail and consider
a modified V-type three-level system (Fig.~\ref{fig:3level}).
It consists of an electric dipole transition which couples the ground state $|1\rangle$
and the excited state $|3\rangle$ as well as a magnetic dipole transition between
$|1\rangle$ and the upper state $|2\rangle$. We assume states $|2\rangle$ and $|3\rangle$
to be energetically degenerate such that the electric ($E$) and magnetic ($B$) field components of
the probe field can couple efficiently to the transitions $|1\rangle-|3\rangle$ and $|1\rangle-|2\rangle$,
respectively.
In order to couple the electric and magnetic dipoles, i.e., to induce a cross-coupling
in the sense of equation (\ref{eq:Pendrychiral})
we add a strong resonant coherent coupling field between the two upper states
$|2\rangle$ and $|3\rangle$ with a Rabi-frequency $\Omega_c$.
Note that the configuration of Fig.~\ref{fig:3level} complies with the requirements
of parity rules.
The magnetization of the system at the probe field frequency is given by the coherence of the
$|1\rangle-|2\rangle$ transition which also gets a contribution induced by the electric field $E$
and likewise the polarization, connected to $|1\rangle-|3\rangle$, is not only induced by $E$ but also by
the magnetic field $B$. Following the discussion above the level scheme of
Fig.~\ref{fig:3level} should therefore show a negative refractive index
without requiring $\Re[\mu]<0$.

Concerning the absorption we note that the main contribution
to the imaginary part of $n$ stems from the permittivity $\varepsilon$.
The radiative decay rate $\gamma_2$ of the magnetic dipole transition 
$|2\rangle-|1\rangle$ is typically a factor $\alpha^2$ smaller than
$\gamma_3$, the decay rate of the electric dipole transition.
As a consequence the strong field $\Omega_c$ couples state $|3\rangle$ strongly
to the {\it metastable} state $|2\rangle$ which, on two-photon resonance, is 
the condition for destructive quantum interference for the imaginary part
of the permittivity, known as electromagnetically induced transparency (EIT) \cite{Fleischhauer05}.
Additionally, for closed loop schemes (Fig.~\ref{fig:3level})
it is known from resonant nonlinear optics based on EIT \cite{Harris-NLO-EIT,Stoicheff-Hakuta},
that the dispersive cross-coupling, which in our case is the magneto-electric cross
coupling, experiences constructive interference.

In essence the coupling of an electric to a magnetic dipole transition
should lead to negative refraction for
significantly smaller densities of scatterers compared to
non-chiral proposals \cite{Oktel04,Thommen-PRL-2006,KastelComment}. We additionally expect the imaginary
part of $\varepsilon$, which represents the major contribution to absorption,
to be strongly suppressed due to quantum interference effects while
simultaneously the cross-coupling terms should be further enhanced.

Though conceptually easy the scheme of Fig.~\ref{fig:3level} has several 
drawbacks which demand a modification of the level structure. 
(i) As stated above the phase of the chirality coefficients $\xi_{EH}$, $\xi_{HE}$
must be adjustable to control the sign of the refractive index and induce $\Re[n]<0$.
As $\Omega_c$ is a {\it dc}-field in the scheme of Fig.~\ref{fig:3level} the phases
of $\xi_{EH}$, $\xi_{HE}$ are solely given by the intrinsic
phase of the transition moments and therefore can not be controlled.
(ii) To suppress the absorption efficiently there must be high-contrast
EIT for the probe field. The critical parameter for this effect is the dephasing rate $\gamma_{21}$
of the coherence $\rho_{21}$ between the two EIT ``ground''-states $|2\rangle$ and $|1\rangle$.
Since state $|2\rangle$ has an energy difference to state $|1\rangle$ on the
order of the probe field frequency, the coherence $\rho_{21}$ is highly susceptible to
additional homogeneous or inhomogeneous 
broadenings which ultimately can destroy EIT.
(iii) The level
structure must be appropriate for media of interest (atoms, molecules, excitons, etc.).
Although the scheme of Fig.~\ref{fig:3level} is not forbidden on
fundamental grounds it is very restricting to require that electric and magnetic transitions be
energetically degenerate while having a common ground state.

\begin{figure}[t]
     \begin{center}
       \includegraphics[width=8cm]{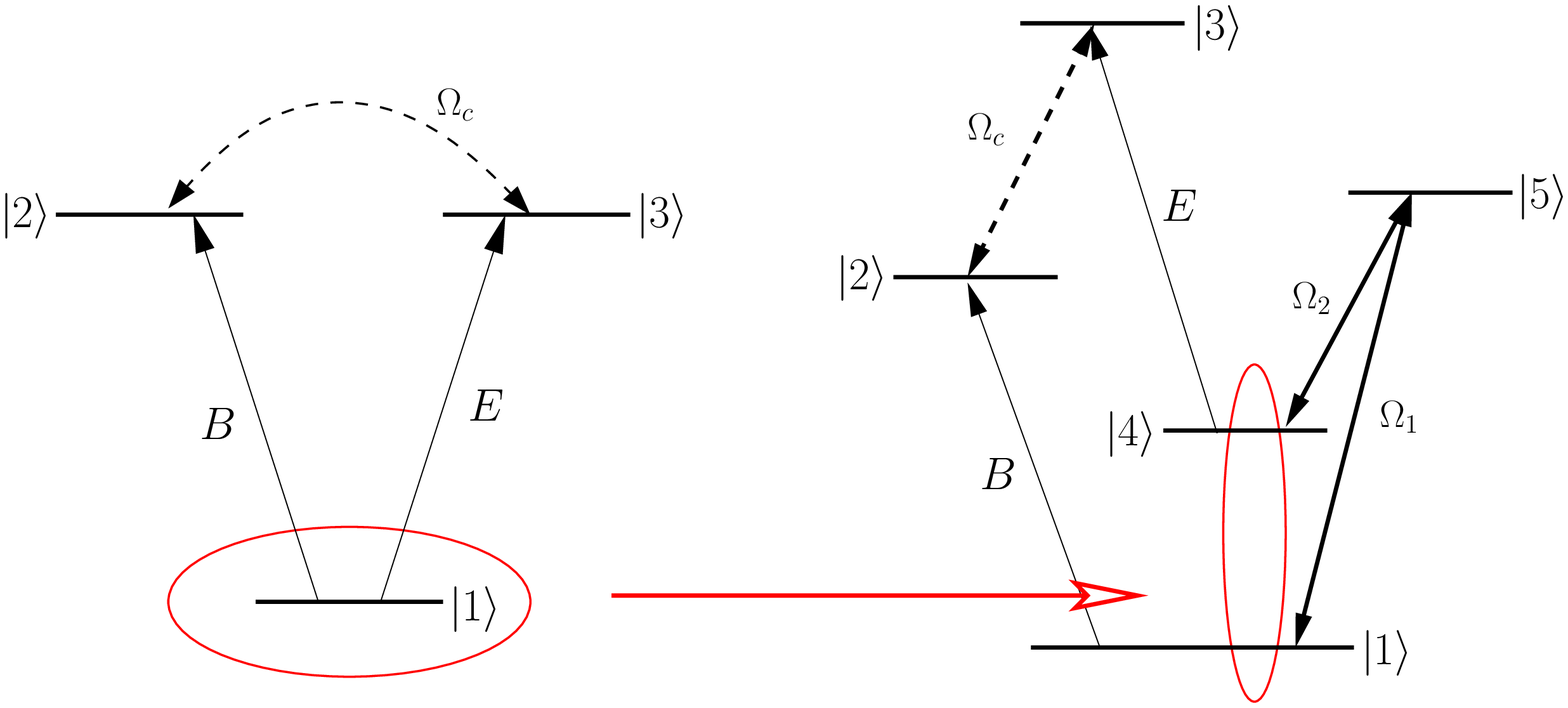}
       \caption{(color online) Modification of the level scheme of Fig.~\ref{fig:3level}. The ground state $|1\rangle$ is substituted by the dark state $|D\rangle=(\Omega_2|1\rangle-\Omega_1|4\rangle)/\sqrt{\Omega_1^2+\Omega_2^2}$ of the 3-level $\Lambda$-type subsystem formed by levels $\{|1\rangle,|4\rangle,|5\rangle\}$.}
       \label{fig:5level}
     \end{center}
\end{figure}

One possible alternative level structure which solves these problems 
is shown in Fig.~\ref{fig:5level}.
The former ground state $|1\rangle$ is now substituted by the dark state $|D\rangle$ of
the 3-level $\Lambda$-type subsystem formed by the new levels $\{|1\rangle,|4\rangle,|5\rangle\}$
as shown in Fig.~\ref{fig:5level}.
$|D\rangle$ is determined by the two coupling field Rabi frequencies $\Omega_1$ and $\Omega_2$:
$|D\rangle=(\Omega_2|1\rangle-\Omega_1|4\rangle)/\sqrt{\Omega_1^2+\Omega_2^2}$.
This simple manipulation indeed solves the above mentioned problems:
The upper states $|2\rangle$ and $|3\rangle$ are no longer degenerate, i.e. the coupling
Rabi frequency $\Omega_c$ is now given by an {\it ac}-field.
By adjustment of $\Omega_c$ the phase of  $\xi_{EH}$ and $\xi_{HE}$
can be controlled. The critical parameter of EIT for this scheme is the dephasing rate $\gamma_{24}$
of the coherence $\rho_{24}$ between states $|2\rangle$ and $|4\rangle$.
By assuming levels $|2\rangle$ and $|4\rangle$ of Fig.~\ref{fig:5level}
to be close to degenerate $\rho_{24}$ can be taken to be insensitive to additional broadenings.
Compared to the 3-level system of Fig.~\ref{fig:3level} the electric and
magnetic transitions here do not share a common state while the transition frequencies are
still degenerate. This leaves much more freedom regarding a realization in real systems.

In the next section we will analyze the 5-level-system in detail.


\section{5-level scheme}
\label{ch:5level}


\subsection{Linear response}
\label{ch:linresp}

\begin{figure}[t]
     \begin{center}
       \includegraphics[width=6cm]{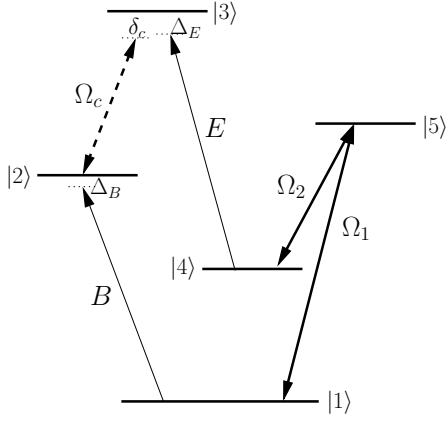}
       \caption{5-level scheme for the implementation of negative refraction via electromagnetically induced cross-coupling. The magnetic dipole transition $|2\rangle-|1\rangle$ and the electric dipole transition $|3\rangle-|4\rangle$ are coupled by $\Omega_c$ to induce chirality. The ``ground''-state of the system is formed by the dark state $|D\rangle=(\Omega_2|1\rangle-\Omega_1|4\rangle)/\sqrt{\Omega_1^2+\Omega_2^2}$ of the subsystem $\{|1\rangle,|4\rangle,|5\rangle\}$.}
       \label{fig:5levelpure}
     \end{center}
\end{figure}

The level scheme in question is shown in greater detail in Fig.~\ref{fig:5levelpure}.
The transition $|2\rangle-|1\rangle$ is magnetic dipole by nature, all other transitions
are electric dipole ones. Note that this complies with the demands of parity.
It is assumed that for reasons of selection rules or absence of resonance
no other transitions than the ones sketched in Fig.~\ref{fig:5levelpure} are allowed.
The Hamiltonian $H = H_\text{atom}-\db\cdot\Eb(t)-\pmb\mu \cdot\Bb(t)$ of the system
can be written explicitly as
\begin{equation}
\begin{split}
H= & \sum_{n=1}^5\hbar\omega^A_n\ketbra{n}{n} \\
&
+\left\{-\frac{1}{2}d_{34}Ee^{-i\omega_pt}\ketbra{3}{4}-\frac{1}{2}\mu_{21}Be^{-i\omega_pt}\ketbra{2}{1}\right. \\
&
-\frac{\hbar}{2}\Omega_1e^{-i\omega_1t}\ketbra{5}{1}-\frac{\hbar}{2}\Omega_2e^{-i\omega_2t}\ketbra{5}{4}\\
& \left. -\frac{\hbar}{2}\Omega_ce^{-i\omega_ct}\ketbra{3}{2}
+\text{h.c.}\right\}.
\label{eq:hamiltonian}
\end{split}
\end{equation}
Here $d_{34}=\bra{3}e\,\rb\cdot\hat{\bf e}_\Eb\ket{4}$ and $\mu_{21}=\bra{2}\pmb\mu\cdot\hat{\bf e}_\Bb\ket{1}$ are 
the electric and  magnetic dipole moments, $E$ and $B$ the electric and magnetic components of
the weak probe field which oscillates at a frequency $\omega_p$. The Rabi frequencies
$\Omega_1$, $\Omega_2$ and $\Omega_c$ belong to strong coupling lasers which
oscillate at frequencies $\omega_1$, $\omega_2$ and $\omega_c$, respectively.
We choose $d_{34}$ and $\mu_{21}$ as well as $\Omega_1$ and $\Omega_2$ to be real whereas the strong
coupling Rabi frequency $\Omega_c$ has to stay complex for the closed loop scheme.

To include losses in our description we solve for the steady state solutions of the Liouville equation
of the density matrix. In doing so we introduce population decay rates $\gamma_i$, $i\in\{1,2,3,4,5\}$.
As we focus on the linear response we treat the probe field amplitudes $E$ and $B$ as weak fields
which allows to neglect the upper state populations $\rho_{33}$ and $\rho_{22}$. In contrast the subsystem
$\{|1\rangle,|4\rangle,|5\rangle\}$ contains strong fields with Rabi frequencies $\Omega_1$ and $\Omega_2$
and should be treated non-perturbatively.

We first solve the 3-level subsystem $\{|1\rangle,|4\rangle,|5\rangle\}$ undisturbed by the probe field.
Under the assumption that $|4\rangle$ is meta-stable and therefore $\gamma_4\ll\gamma_5$
holds, the exact solution is:
\begin{equation}
\begin{split}
\rho_{11}^{(0)}=\frac{|\Omega_2|^2}{|\Omega_1|^2+|\Omega_2|^2} & \qquad
\rho_{44}^{(0)}=\frac{|\Omega_1|^2}{|\Omega_1|^2+|\Omega_2|^2} \\
\tilde\rho_{41}^{(0)}=-\frac{\Omega_1\Omega_2}{|\Omega_1|^2+|\Omega_2|^2}
\qquad & \rho_{55}^{(0)}=\tilde\rho_{51}^{(0)}=\tilde\rho_{54}^{(0)}=0.
\end{split}
\end{equation}
This solution for the $\Lambda$-type subsystem indeed corresponds to the pure state $|D\rangle=(\Omega_2|1\rangle-\Omega_1|4\rangle)/\sqrt{\Omega_1^2+\Omega_2^2}$ via
$\rho^{\rm subsys.}=\ketbra{D}{D}$.
Note that density matrix components with tildas denote slowly varying quantities.

We proceed by solving for the polarizabilities of the complete 5-level-scheme
up to first order in the probe field amplitudes $E$ and $B$. Since
the induced Polarization $P$ is proportional to the coherence
of the electric dipole transition $\tilde\rho_{34}$ whereas the induced Magnetization $M$ is 
proportional to the density matrix element $\tilde\rho_{21}$ we arrive at
\begin{equation}
\begin{split}
P = & \varrho\, d_{34} \tilde\rho_{34}:=\varrho\,\alpha^{EE}E+\varrho\,\alpha^{EB}B \\
M = & \varrho\,\mu_{21} \tilde\rho_{21}:=\varrho\,\alpha^{BE}E+\varrho\,\alpha^{BB}B.
\label{eq:alphas}
\end{split}
\end{equation}
Here $\varrho$ is the number density of scatterers, $\alpha^{EE}$, $\alpha^{EB}$,
$\alpha^{BE}$ and $\alpha^{BB}$ are the direct and the cross-coupling polarizabilities.
They are given by:
\begin{equation}
\alpha^{EE}=\frac{i}{2\hbar}\frac{d_{34}^2\rho_{44}^{(0)}(\gamma_{42}+i(\Delta_E-\delta_c))}{(\gamma_{42}+i(\Delta_E-\delta_c))(\gamma_{34}+i\Delta_E)+|\Omega_c|^2/4},
\label{eq:polEE}
\end{equation}
\begin{equation}
\alpha^{BB}=\frac{i}{2\hbar}\frac{\mu_{21}^2\rho_{11}^{(0)}(\gamma_{31}+i(\Delta_B+\delta_c))}{(\gamma_{31}+i(\Delta_B+\delta_c))(\gamma_{21}+i\Delta_B)+|\Omega_c|^2/4}
\label{eq:polBB}
\end{equation}
as well as
\begin{equation}
\alpha^{EB}=-\frac{1}{4\hbar}\frac{d_{34}\mu_{21}\tilde\rho_{41}^{(0)}\Omega_c}{(\gamma_{42}+i(\Delta_E-\delta_c))(\gamma_{34}+i\Delta_E)+|\Omega_c|^2/4},
\label{eq:polEB}
\end{equation}
\begin{equation}
\alpha^{BE}=-\frac{1}{4\hbar}\frac{d_{34}\mu_{21}\tilde\rho_{41}^{(0)}\Omega_c^\ast}{(\gamma_{31}+i(\Delta_B+\delta_c))(\gamma_{21}+i\Delta_B)+|\Omega_c|^2/4}.
\label{eq:polBE}
\end{equation}
Here the definitions $\gamma_{ij}=(\gamma_i+\gamma_j)/2$, $\Delta_E=\omega_{34}-\omega_p$, 
$\Delta_B=\omega_{21}-\omega_p$ and $\delta_c=\omega_{32}-\omega_c$ apply,
with $\omega_{\mu\nu}=\omega_\mu^A-\omega_\nu^A$ being the
transition frequencies between levels $|\mu\rangle$ and $|\nu\rangle$.
Note that these solutions are only valid for $\Delta_E=\Delta_B+\delta_c$ which
corresponds to the resonance condition
\begin{equation}
\omega_c=\omega_1-\omega_2
\end{equation}
which ensures the total frequency in the closed loop scheme to sum up to zero.

In order to visualize the polarizabilities we set the magnetic dipole decay rate to
a typical radiative value for optical frequencies \cite{Cowan} of
$\gamma_2=1$kHz, the electric dipole decay rates correspondingly to $\gamma_3=\gamma_5=137^2\gamma_2$
and $\gamma_1=\gamma_4=0$ for the (meta-)stable states $|1\rangle$ and $|4\rangle$.
The electric and magnetic dipole matrix elements $d_{34}$ and
$\mu_{21}$ are determined from the radiative decay rates via the Wigner-Weisskopf result \cite{Louisell}
$d_{34}(\mu_{21})=\sqrt{3\gamma_3(\gamma_2)\hbar c^3/(4\omega^3)}$
for an optical frequency corresponding to $\lambda=600$ nm.
The Rabi frequencies of the $\Lambda$-type subsystem attain real values
$\Omega_1=\Omega_2=10^2\gamma_2$ while the coupling Rabi frequency $\Omega_c$
is chosen complex, $\Omega_c=|\Omega_c| e^{i\phi}$.
We furthermore specialize to $\delta_c=0$ which implies $\Delta_E=\Delta_B$. In order to
have increasing photon energy from left to right  in figures presented here all following spectra are plotted
as a function of $\Delta=-\Delta_E=-\Delta_B$.

\begin{figure}[t]
     \begin{center}
       \includegraphics[width=8cm]{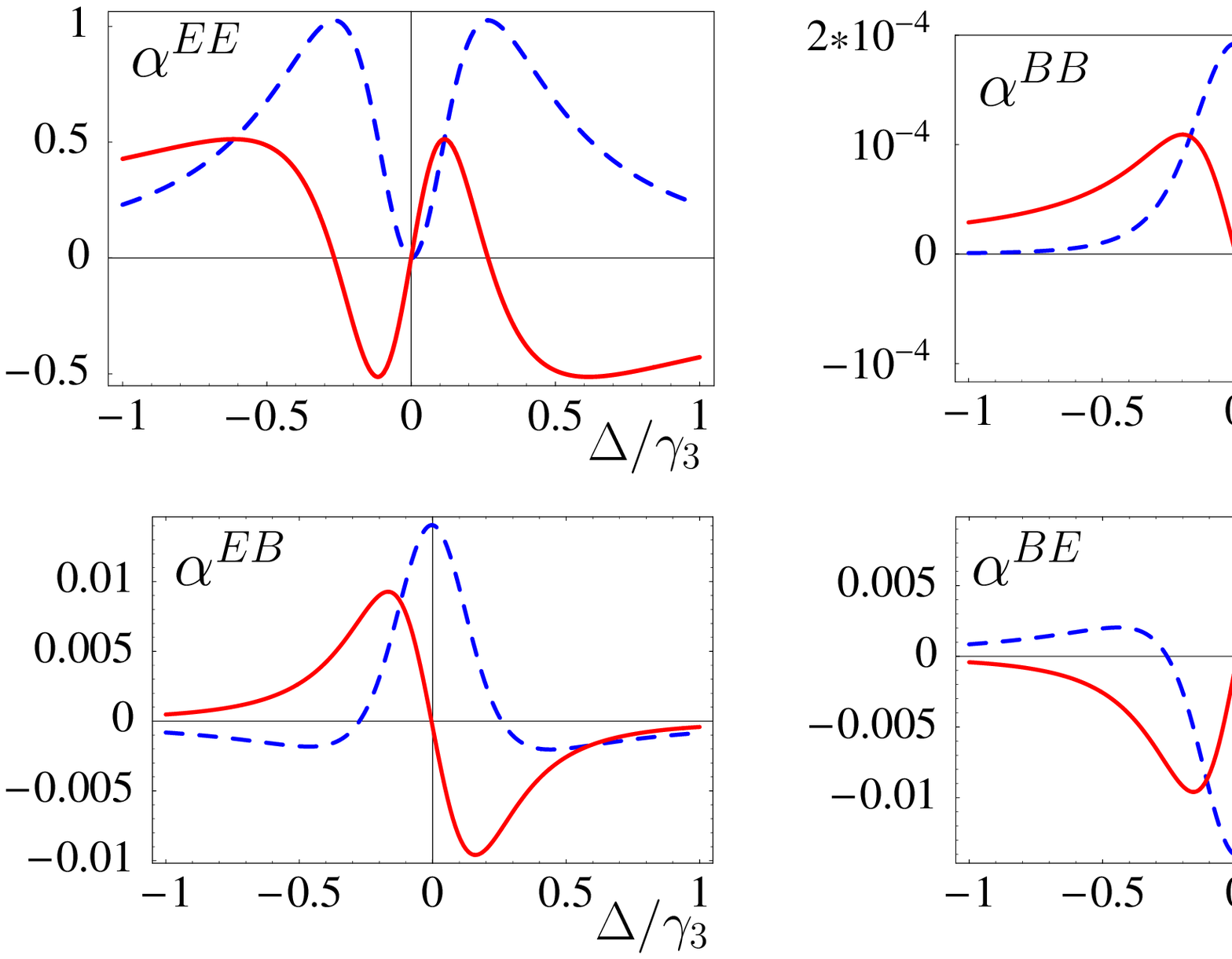}
       \caption{(color online) Real (solid lines) and imaginary (dashed lines) parts of the electric ($\alpha^{EE}$) and magnetic ($\alpha^{BB}$) polarizabilities as well as the cross-coupling parameters ($\alpha^{EB}$, $\alpha^{BE}$) for arbitrary but the same units for $\Omega_{c}=10^4\gamma_2e^{i\pi/2}$.}
       \label{fig:Coup_noBroad}
     \end{center}
\end{figure}
For $\Omega_c=0$ the cross-coupling coefficients $\alpha^{EB}$ and $\alpha^{BE}$
vanish exactly whereas the electric as well as the magnetic polarizability both
show a simple Lorentzian resonance. 
Introducing a non-zero coupling strength $\Omega_c$ changes the response
dramatically. As shown in Fig.~\ref{fig:Coup_noBroad}
for $\Omega_c=10^4\gamma_2e^{i\pi/2}$ the electric polarizability $\alpha^{EE}$ 
indeed displays electromagnetically induced transparency (EIT).
The value of $|\Omega_c|$ is optimized as will be discussed in section \ref{ch:nonrad}.
As long as the coupling field $\Omega_c$ is
present Im$[\alpha^{EE}]$ on resonance is proportional to the decoherence rate $\gamma_{42}$ of
the two EIT ground states $|4\rangle$ and $|2\rangle$ which can become very small. Thus the 
prominent feature of EIT emerges: suppression of absorption.
In contrast the magnetic polarizability $\alpha^{BB}$ still shows an ordinary 
Lorentzian resonance since $\gamma_{31}$ is always large. 
Due to the coupling to the strong electric dipole transition
the magnetic resonance is broadened as compared to its radiative
linewidth which is accompanied by a significant decrease of the magnetic response
susceptibility on resonance.

For a non-vanishing $\Omega_c$ the two cross-couplings $\alpha^{EB}$ and $\alpha^{BE}$
show strongly peaked spectra. Note that the phase $\phi=\pi/2$ has been chosen such that
on resonance $\alpha^{EB}=-\alpha^{BE}\propto i$ holds, as demanded in section \ref{ch:concepts}.

Note furthermore that we verified numerically that all polarizabilities and cross-coupling 
terms are causal and thus fulfill Kramers-Kronig relations.


\subsection{Limits of linear response theory}
\label{ch:saturation}


For radiatively broadened systems
$\gamma_{21}^{(B)}\approx\alpha^2\gamma_{21}^{(E)}$ holds. Thus
magnetic transitions saturate at much lower field amplitudes than corresponding electric dipole
transitions. For this reason we have to analyze the saturation behavior of the system. 
\begin{figure}[t]
     \begin{center}
       \includegraphics[width=8cm]{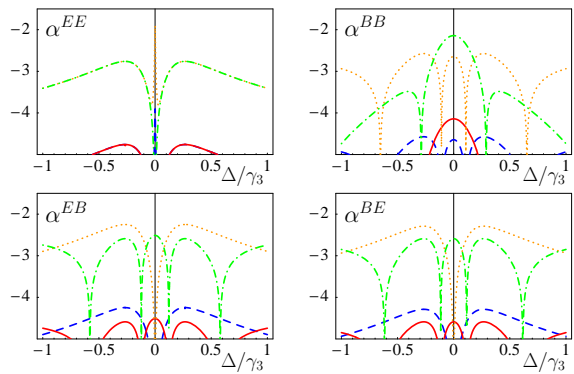}
       \caption{(color online) Deviation (\ref{eq:deviation}) of real and imaginary parts of the exact polarizabilities $\alpha^{IJ}(E,B)$ compared to the linear response results for 2 different probe field Rabi frequencies: $\Omega_E=137\,\Omega_B=\gamma_2$ (real: solid lines, imaginary: dashed lines) and $\Omega_E=137\,\Omega_B=10\cdot\gamma_2$ (real: dash-dotted lines, imaginary: dotted lines).}
       \label{fig:Saturation}
     \end{center}
\end{figure}

To rule out saturation effects and validate the use of linear response theory
we solve the Liouvillian equation for the 5-level system of Fig.~\ref{fig:5levelpure}
to all orders in the electric and magnetic field amplitudes $E$ and $B$ which
can only be done numerically.
We determine the polarizabilities $\alpha^{EE}(E,B)$, $\alpha^{BE}(E,B)$,
$\alpha^{BB}(E,B)$ and $\alpha^{EB}(E,B)$ from the numerically accessible density matrix elements
$\tilde\rho_{34}$ and $\tilde\rho_{21}$ (see \ref{app:B}) for the worst case scenario of
purely radiatively broadened transitions and compare to the results of the linear response theory.
Fig.~\ref{fig:Saturation} shows the deviation of the exact solution to the linear response results
\begin{equation}
\log{\left|1-\frac{\alpha^{IJ}(E,B)}{\alpha^{IJ}}\right|}
\label{eq:deviation}
\end{equation}
for all polarizabilities.
The solid lines correspond to $\Omega_E=137\,\Omega_B=\gamma_2$,
the dashed lines to $\Omega_E=137\,\Omega_B=10\cdot\gamma_2$.
Here $\Omega_E=d_{34}E/\hbar$ and $\Omega_B=\mu_{21}B/\hbar$
denote the electric and magnetic probe field Rabi frequencies for the 5-level system.

For a radiatively broadened electric dipole 2-level atom
we can estimate the probe-field Rabi frequency required to lead to a 
$1\%$ upper state population to be $\tilde\Omega_E=
\sqrt{0.01} \gamma_3 \approx 13.7\gamma_2$. 
>From Fig.~\ref{fig:Saturation} we note that even for
$\Omega_E=137\,\Omega_B=10\times\gamma_2$, i.e., the same order of magnitude as $\tilde\Omega_E$,
the deviation of the exact result from the spectrum obtained in linear response approximation
never exceeds $10^{-2}$. As a result we conclude
that the 5-level scheme is not significantly more sensitive to saturation effects
than any ordinary electric dipole transition.

This behavior is a result of the coupling Rabi frequency $\Omega_c$ due to which
the $|2\rangle-|1\rangle$ transition experiences an additional broadening which makes it
less susceptible to saturation.


\subsection{Non-radiative broadenings}
\label{ch:nonrad}


As noted in section \ref{ch:introduction} additional broadenings are essential for
the description of spectral properties as soon as magnetic transitions are involved.

To add an additional homogeneous dephasing rate we formally add to every static
detuning $\Delta_E$, $\Delta_B$, $\delta_c$ an extra random
term $x_E$, $x_B$, $x_c$ (e.g. $\Delta_E\longrightarrow\Delta_E+x_E$)
which has to be convoluted with a Lorentzian distribution
\begin{equation}
L(x)=\frac{1}{\pi}\frac{\gamma_p}{\gamma_p^2+x^2}.
\end{equation}
For, e.g., $\alpha^{EE}$ this amounts to
\begin{equation}
\tilde\alpha^{EE}=\int\! dx_Edx_Bdx_c\alpha^{EE}(x_E,x_B,x_c)L(x_E)L(x_B)L(x_c).
\end{equation}
Levels $\ket{2}$ and $\ket{4}$ are
approximately degenerate and thus assumed to experience correlated
phase fluctuations. As a consequence  $\gamma_{42}$, which is relevant for EIT, remains unchanged while
the same width $\gamma_p$ applies for both $\Delta_E$ and $\delta_c$.
For simplicity we choose the same width for $\Delta_B$.
The convolution integral can be solved analytically which results 
in the substitution of the off-diagonal decay rates $\gamma_{ij},i\neq j$ 
(\ref{eq:polEE})-(\ref{eq:polBE}) according to
\begin{equation}
\begin{array}{ll}
\gamma_{42} \longrightarrow \gamma_{42} & \qquad \gamma_{21} \longrightarrow \gamma_{21}+\gamma_p \\
\gamma_{34} \longrightarrow \gamma_{34}+\gamma_p & \qquad \gamma_{31} \longrightarrow \gamma_{31}+2\gamma_p.
\end{array}
\label{eq:broadenings}
\end{equation}
In contrast $\gamma_{21}$ and $\gamma_{34}$ encounter a broadening
$\gamma_p$, likewise $\gamma_{31}$ experiences a broadening $2\gamma_p$.

\begin{figure}[t]
     \begin{center}
       \includegraphics[width=8cm]{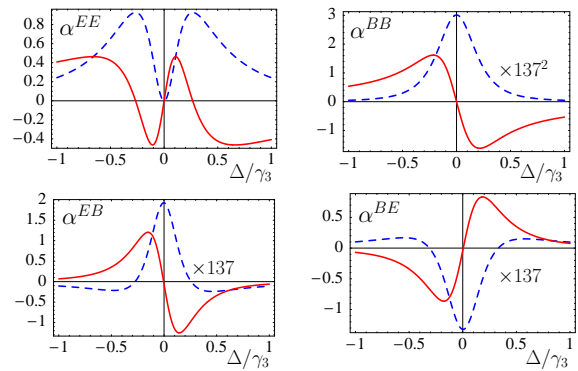}
       \caption{(color online) Real (solid lines) and imaginary (dashed lines) parts of the electric ($\alpha^{EE}$) and magnetic ($\alpha^{BB}$) polarizabilities as well as the cross-coupling parameters ($\alpha^{EB}$, $\alpha^{BE}$) for arbitrary but the same units for $\Omega_c=10^4\gamma_2e^{i\pi/2}$. In contrast to Fig.~\ref{fig:Coup_noBroad} additional homogeneous broadenings according to \eq{eq:broadenings} with $\gamma_p=10^3\gamma_2$ apply.}
       \label{fig:Coup_Broad}
     \end{center}
\end{figure}

We choose the value $\gamma_p=10^3\gamma_2$ which is typical for rare-earth doped
crystals at cryogenic temperatures. 
For a given broadening $\gamma_p$ the cross-coupling terms
reach a maximum for the coupling Rabi frequency attaining the optimal values 
$137\gamma_2$ and $1370\sqrt{20}\gamma_2$, respectively.
Fig.~\ref{fig:Coup_Broad} shows the polarizabilities for an intermediate value of
$\left|\Omega_c\right|=10^4\gamma_2$. Since $\gamma_{42}$ remains unbroadened
the electric polarizability $\alpha^{EE}$ still shows EIT while the spectrum of $\alpha^{BB}$
shows a simple but broadened resonance with $\alpha^{EE}\approx 137^2\alpha^{BB}$.
Similarly $\alpha^{EE}\approx 137\alpha^{EB}$ and
$\alpha^{EE}\approx 137\alpha^{BE}$ hold approximately.

To incorporate the effect of an inhomogeneous Doppler broadening mechanism on the
spectrum the same formalism as for the homogeneous case, but with a Gaussian
instead of a Lorentzian distribution, can be used.


\subsection{Local field correction}


So far we have dealt with a single individual radiator (e.g., atom) responding to the locally acting fields.
For large responses these local fields are known to differ from the averaged Maxwell field.
We therefore have to correct the results of section \ref{ch:linresp} by the use of
Clausius-Mossotti type local field corrections. Note that due to the cross-coupling 
the influence of the magnetic properties is enhanced by a factor of approximately $137$. 
We therefore also include magnetic local field corrections in the treatment.

To add local field corrections to the response the fields $E$ and $B$
\eq{eq:alphas} are interpreted as local or microscopic ones:
\begin{equation}
\begin{split}
P = & \varrho\,\alpha^{EE}E^{\rm micro}+\varrho\,\alpha^{EB}B^{\rm micro} \\
M = & \varrho\,\alpha^{BE}E^{\rm micro}+\varrho\,\alpha^{BB}B^{\rm micro}
\label{eq:alphasmicro}
\end{split}
\end{equation}
($\varrho$: number density of scatterers).
The relations between the local and the corresponding
macroscopic field amplitudes can be obtained from 
phenomenological considerations \cite{Jackson,Cook} which read
\begin{equation}
  E^{\rm micro}= E+\frac{4\pi}{3}P, \qquad
  H^{\rm micro}= H+\frac{4\pi}{3}M
\label{eq:micromacro}
\end{equation}
for the electric and the magnetic field, respectively.
Note that we need to replace $B$ by $H$ to find the permittivity $\varepsilon$, permeability $\mu$, and
coefficients $\xi_{EH}$ and $\xi_{HE}$ of equation (\ref{eq:Pendrychiral})
in terms of the polarizabilities of equations (\ref{eq:polEE}) - (\ref{eq:polBE}).
This can be done most easily for the local microscopic fields for which $B^{\rm micro}=H^{\rm micro}$
holds \cite{Cook}.
Solving \eq{eq:alphasmicro} together with (\ref{eq:micromacro}) for $P$ and $M$ in
terms of the macroscopic field amplitudes $E$ and $H$ yields
\begin{equation}
\begin{split}
\varepsilon = & 1+ 4\pi\frac{\varrho }{{\mathcal L}^\text{loc}} \times\\
& \left\{\alpha^{EE} +\frac{4\pi}{3}\varrho\Bigl(\alpha^{EB}\alpha^{BE} -\alpha^{EE}\alpha^{BB}\Bigr)\right\}, \\
\mu = & 1 + 4\pi\frac{\varrho}{{\mathcal L}^\text{loc}} \times\\
& \left\{\alpha^{BB} +\frac{4\pi}{3}\varrho
\Bigl(\alpha^{EB}\alpha^{BE}-\alpha^{EE}\alpha^{BB}\Bigr)\right\},
\label{eq:responsefunctions}
\end{split}
\end{equation}
\begin{equation}
\begin{split}
\xi_{EH} = & 4\pi\frac{\varrho}{{\mathcal L}^\text{loc}}\alpha^{EB}, \\
\xi_{HE} = & 4\pi\frac{\varrho}{{\mathcal L}^\text{loc}}\alpha^{BE},
\end{split}
\label{eq:chiralities}
\end{equation}
with the denominator
\begin{equation*}
\begin{split}
{\mathcal L}^\text{loc} = & 1-\frac{4\pi}{3}\varrho\,\alpha^{EE}-\frac{4\pi}{3}\varrho\,\alpha^{BB}
\\
& -\left(\frac{4\pi}{3}\right)^2 \varrho^2 \Bigl(\alpha^{EB}\alpha^{BE}-\alpha^{EE}\alpha^{BB}\Bigr).
\end{split}
\end{equation*}
Note that for media without a magneto-electric cross-coupling
a rigorous microscopic derivation \cite{Kaestel07}
validates the phenomenological procedure adopted here.


\subsection{Negative refraction with low absorption}
\label{ch:negrefract}


With the permittivity $\varepsilon$ and the permeability $\mu$ given by
\eq{eq:responsefunctions} and the parameters 
$\xi_{EH}$ and $\xi_{HE}$ [\eq{eq:chiralities}]
we determine the index of refraction from \eq{eq:refind}.
\begin{figure}[t]\centering
\subfigure[$\varrho=5\cdot 10^{14}\text{cm}^{-3}$]{\includegraphics[width=3.8cm]{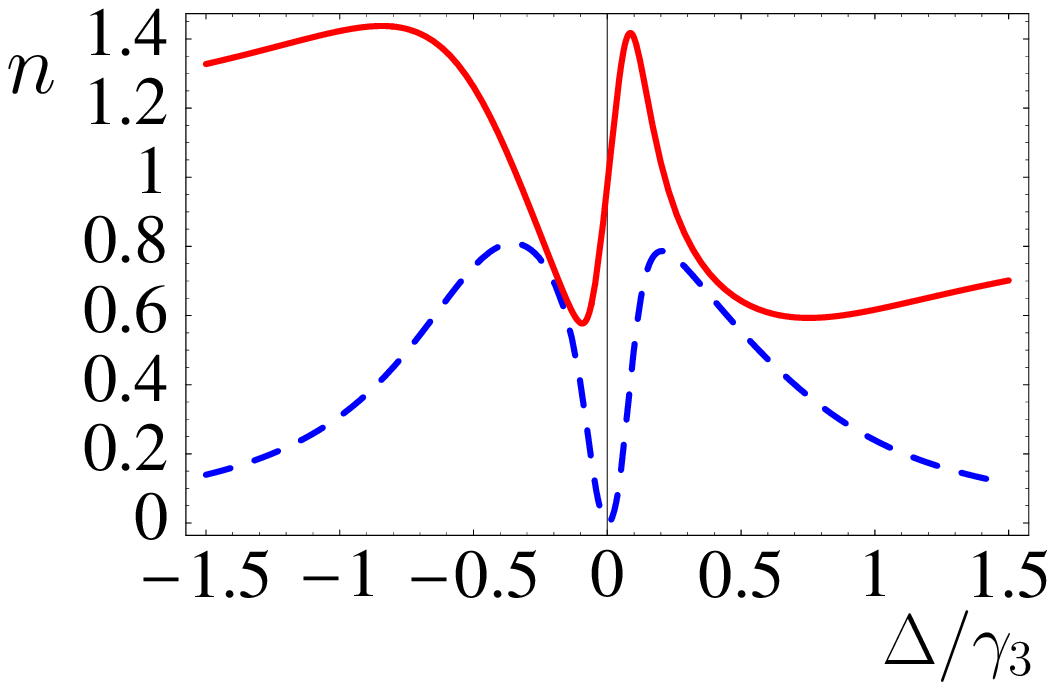}}
\hspace{5mm}
\subfigure[$\varrho=5\cdot 10^{16}\text{cm}^{-3}$]{\includegraphics[width=3.8cm]{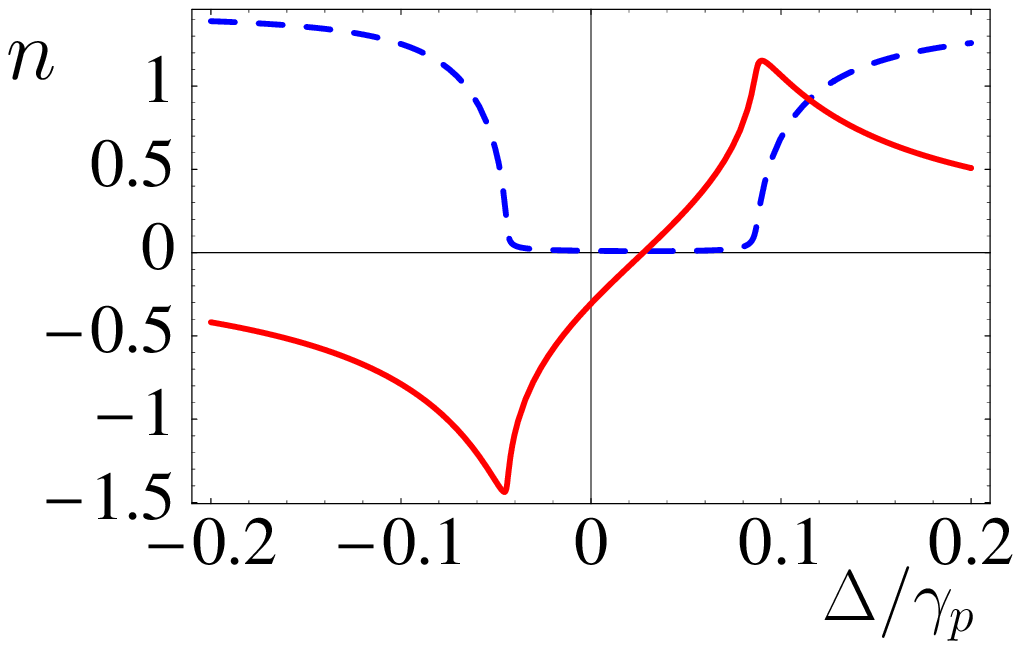}}
\caption{(color online) Real (solid lines) and imaginary (dashed lines) parts of the refractive index, including local field effects for two different densities.}
\label{fig:refractindex1416}
\end{figure}
As an example, Fig.~\ref{fig:refractindex1416}(a) shows the calculated real and imaginary
parts of the refractive index as a function of the probe field detuning $\Delta$ for a density
of $\varrho=5\cdot 10^{14}\text{cm}^{-3}$ and otherwise using the parameter
values defined in sections \ref{ch:linresp} and \ref{ch:nonrad}. 
The shape of the spectrum is governed by the permittivity $\varepsilon$
with the prominent features of EIT: suppression of absorption and steep slope of
the dispersion on resonance. Clearly for this density there is no negative refraction yet.

In Fig.~\ref{fig:refractindex1416}(b) the spectrum of $n$ for an increased density of
$\varrho=5\cdot 10^{16}\text{cm}^{-3}$ is shown. Note that in contrast to Fig.~\ref{fig:refractindex1416}(a) the frequency axis is scaled in units of the broadening $\gamma_p$
as local field effects start to influence the shape of the spectral line at this density.
We find substantial negative refraction and minimal absorption for this density.
The density is about a factor $10^2$ smaller than the density needed without
taking chirality into account \cite{Oktel04}.

\begin{figure}[t]\centering
\subfigure[]{\includegraphics[width=3.7cm]{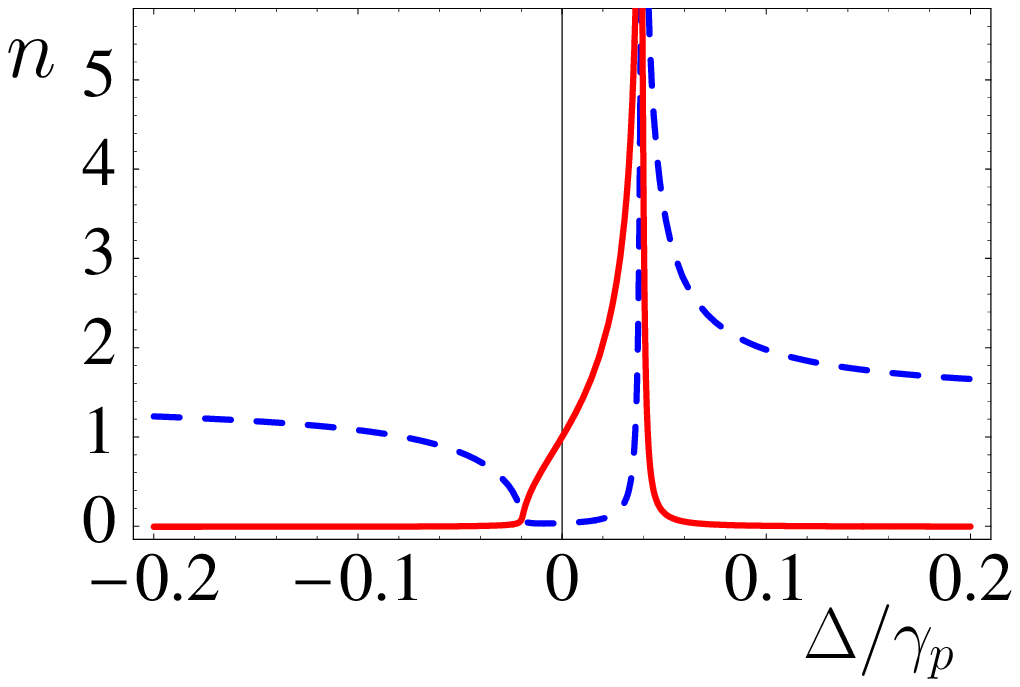}}
\hspace{5mm}
\subfigure[]{\includegraphics[width=4.1cm]{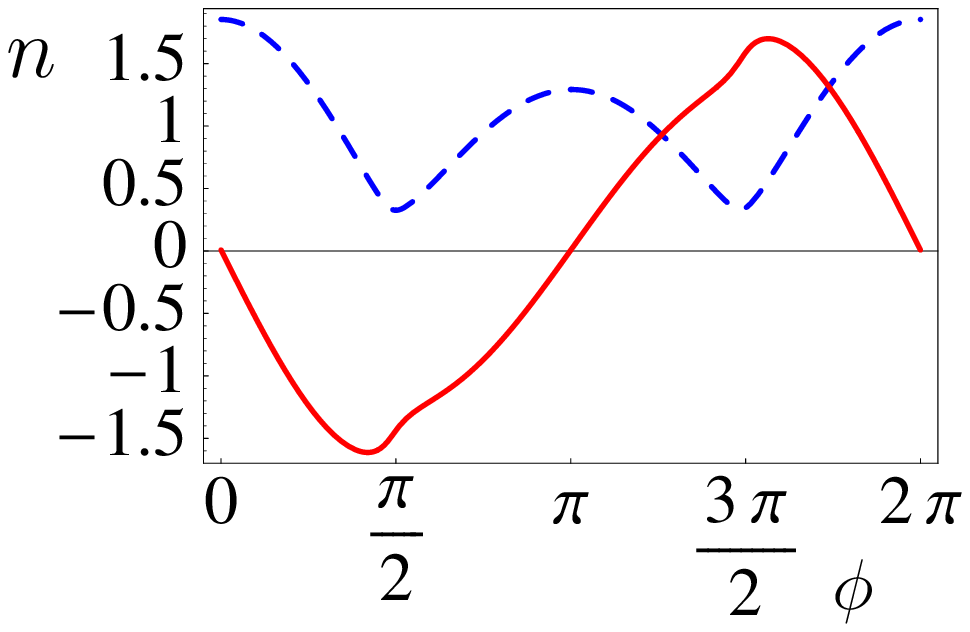}}
\caption{(color online) (a) Real (solid line) and imaginary (dashed line) parts of the refractive index for $\varrho=5\cdot 10^{16}\text{cm}^{-3}$. Compared to Fig.~\ref{fig:refractindex1416}(b) here $\tilde\rho_{41}^{(0)}=0$ applies. (b) Real (solid line) and imaginary (dashed line) parts of the refractive index as a function of the phase $\phi$ of the coupling Rabi frequency $\Omega_c$.}
\label{fig:refractindex16NC_Phase}
\end{figure}

In order to validate that the negative index results from the cross-coupling we compare
the spectrum of Fig.~\ref{fig:refractindex1416}(b) to a non-chiral version.
As setting $\Omega_c=0$ influences the permittivity and permeability as well, we set
$\tilde\rho_{41}^{(0)}=0$ by hand such that the cross-couplings vanish identically.
The resulting index of refraction is shown in Fig.~\ref{fig:refractindex16NC_Phase}(a)
for a density $\varrho=5\cdot 10^{16}\text{cm}^{-3}$. We find that without cross-coupling
no negative refraction occurs. Thus the negative index at this density is clearly
a consequence of the cross-coupling.

Following the qualitative discussion in section \ref{ch:concepts} we have set the phase $\phi$
of the coupling Rabi frequency $\Omega_c$ to $\phi=\pi/2$.
Fig.~\ref{fig:refractindex16NC_Phase}(b) shows the dependence of the refractive index on $\phi$
taken at the spectral position $\Delta=-0,045\gamma_p$ at which $\Re[n]$ reaches its minimum for
$\varrho=5\cdot 10^{16}\text{cm}^{-3}$. As expected the refractive index is strongly phase
dependent. A change of the phase by $\delta\phi=\pi$ for example reverses the influence of
the cross-coupling and gives a positive index of refraction $\Re[n]>0$.
Note that the symmetry $\Re[n(\phi)]=-\Re[n(2\pi-\phi)]$ is coincidental since for the chosen
parameters $\varepsilon\approx 0$.

By increasing the density of scatterers $\varrho$ further the optical response of the medium increases.
\begin{figure}[t]\centering
 \subfigure[]{\includegraphics[width=3.9cm]{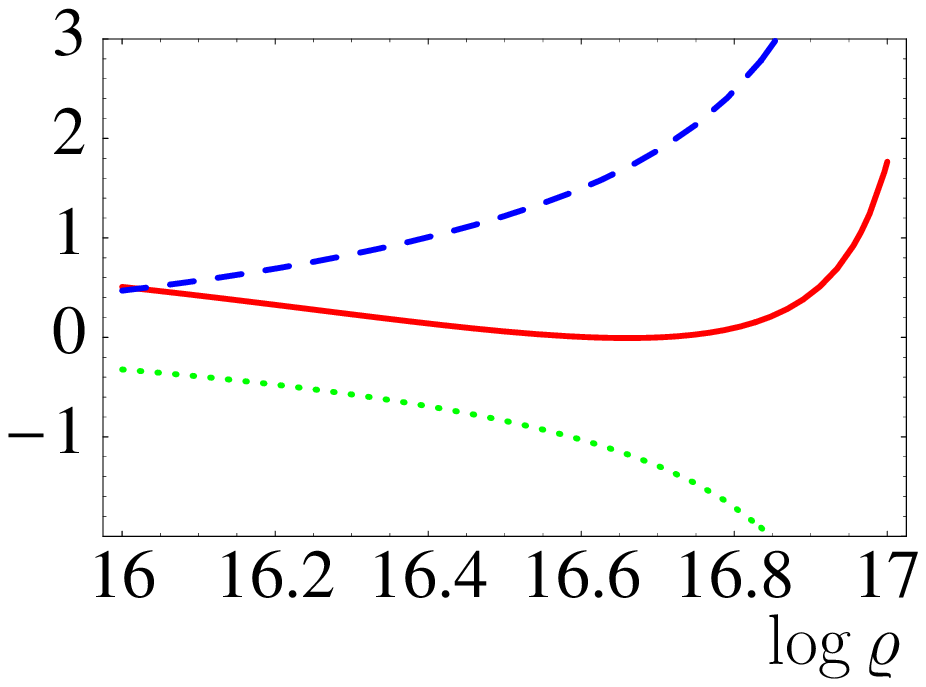}}
\hspace{5mm}
 \subfigure[]{\includegraphics[width=3.8cm]{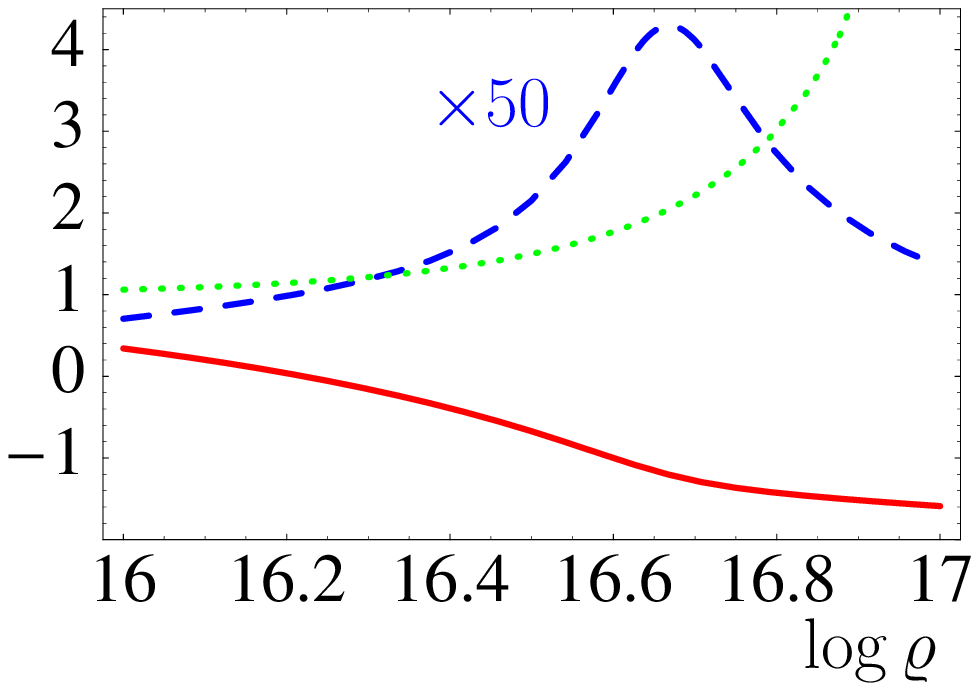}}
   \caption{(color online) Real part of (a) the permittivity $\varepsilon$ (solid line) as well as the imaginary parts of $\xi_{EH}$ (dashed line) and $\xi_{HE}$ (dotted line) and (b) real (solid line) and imaginary (dashed line, $\times 50$) parts of the refractive index, as well as the real part of the permeability (dotted line) as a function of the logarithm of the density $\log \varrho$.}
\label{fig:dens}
\end{figure}
For a spectral position slightly below resonance ($\Delta=-0,045\gamma_p$) 
where negative refraction is obtained most effectively we show $\Re[\varepsilon]$ as
well as $\Im[\xi_{EH}]$ and $\Im[\xi_{HE}]$ as functions of the density $\varrho$
[Fig.~\ref{fig:dens}(a)]. Due to local field corrections the permittivity is of the same order
of magnitude as the cross-coupling terms. The imaginary parts of the parameters $\xi_{EH}$ and
$\xi_{HE}$ increase strongly with opposite signs causing the refractive index to become negative.
The corresponding density dependence of the refractive index is shown in Fig.~\ref{fig:dens}(b).
We find that $\Re[n]$ becomes negative while the absorption $\Im[n]$ stays small
(note that in Fig.~\ref{fig:dens}(b) $\Im[n]$ is amplified by a factor of 50).
Additionally $\Re[\mu]$ is positive and becomes larger for increasing density
as a consequence of operating on the red detuned side of the resonance ($\Delta<0$).

\begin{figure}[t]
     \begin{center}
        \includegraphics[width=6.5cm]{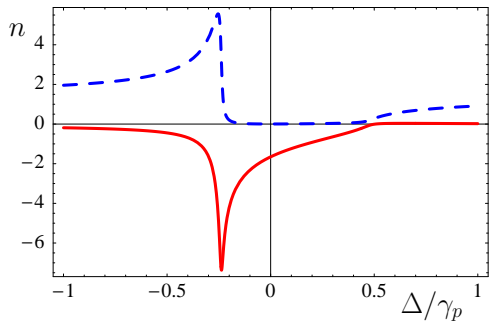}
      \caption{(color online) Real (solid line) and imaginary (dashed line) parts of the refractive index, including local field effects for a density of $\varrho=5\cdot 10^{17}\text{cm}^{-3}$.}
       \label{fig:refractindex17}
     \end{center}
\end{figure}

As an example for higher densities the spectrum of $n$ is shown for 
$\varrho=5\cdot 10^{17}\text{cm}^{-3}$ in Fig.~\ref{fig:refractindex17}.
Compared to the case of $\varrho=5\cdot 10^{16}\text{cm}^{-3}$ [Fig.~\ref{fig:refractindex1416}(b)]
$\Im[n]$ did not change much qualitatively while $\Re[n]$ reaches larger negative values.

Remarkably, in Fig.~\ref{fig:dens}(b) we find that the absorption $\Im[n]$
reaches a maximum and then decreases with increasing density of scatterers.
This peculiar behavior is due to local field effects which invariably get important at such
high values of the response. As a consequence the spectral band with
minimal absorption broadens with increasing density due to local field effects.
Hence the chosen spectral position moves from the tail of the band edge to the
middle of the minimal absorption band.

\begin{figure}[t]
     \begin{center}
        \includegraphics[width=6.5cm]{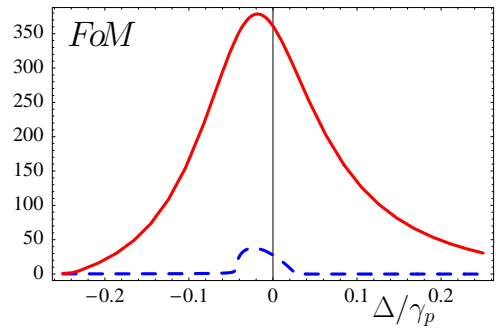}
     \caption{(color online) The Figure of Merit $\fom$ for densities $\varrho=5\cdot 10^{17}\text{cm}^{-3}$ (solid line) and $\varrho=5\cdot 10^{16}\text{cm}^{-3}$ (dashed line) as a function of the detuning $\Delta$.}
       \label{fig:FOM}
     \end{center}
\end{figure}

As a consequence of the low absorption and corresponding increasing values
of $\Re[n]$ the $\fom$ continues to increase with density and reaches rather large values.
In Fig.~\ref{fig:FOM}
we show the $\fom$ as a function of $\Delta$ for $\varrho=5\cdot 10^{16}\text{cm}^{-3}$
and $\varrho=5\cdot 10^{17}\text{cm}^{-3}$.
While the $\fom$~reaches for $\varrho=5\cdot 10^{16}\text{cm}^{-3}$ values of $\approx 35$
it climbs for $\varrho=5\cdot 10^{17}\text{cm}^{-3}$ up to $\fom\approx 350$.
These results should be contrasted to previous
theoretical proposals and experiments on negative refraction in the optical
regime, for which the refraction/absorption ratio is typically on the
order of unity.


\section{Tunability}
\label{ch:tunable}


As first noted by 
Smith {\it et al.} \cite{Smith03} and Merlin \cite{Merlin} sub-wavelength imaging 
using a flat lens of thickness $d$ requires not
only some negative refractive index but an extreme control of the absolute value of $n$.
For an intended resolution $\Delta x$ the accuracy with which the value $n=-1$ (assuming a vacuum
environment) has to be met is given by 
\begin{equation}
|\Delta n|= \exp\left\{-\frac{2\pi d}{\Delta x}\right\}.
\label{eq:finetuning}
\end{equation}
For a metamaterial with $\Re[n]<0$ \eq{eq:finetuning} presents a considerable obstacle
for the operation of a superlens approaching far field distances ($d\gg \lambda$) as it demands an extreme
fine-tuning of the refractive index in order to achieve a resolution beyond the diffraction limit.

\begin{figure}[t]
     \begin{center}
        \includegraphics[width=6.5cm]{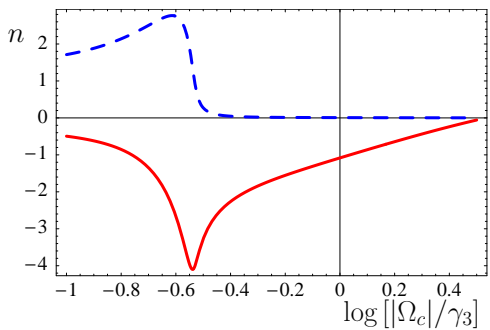}
        \caption{(color online) Real (solid line) and imaginary (dashed line) parts of the refractive index as a function of the coupling field Rabi-frequency $|\Omega_c|$ relative to the radiative decay rate $\gamma_3$, for $\varrho=1,56\cdot10^{17}$cm$^{-3}$.}
       \label{fig:tunability}
     \end{center}
\end{figure}

Our scheme allows to achieve such a fine tuning. In Fig.~\ref{fig:tunability} we show
the real and imaginary parts of $n$ as a function of $\log[|\Omega_{c}|/\gamma_3]$
for a density of $\varrho=1,56\cdot10^{17}$cm$^{-3}$.
As the coupling Rabi frequency approaches $\gamma_3$ we find small $\Im[n]$ while $\Re[n]$ 
attains negative values.
The dispersion then changes only slightly with a small slope and values around
$n=-1$. Therefore the refractive index can be fine tuned by relatively
coarse adjustments of the strength of the coupling field $\Omega_c$.
Note that the value and slope of $n$ for $|\Omega_{c}|\approx\gamma_3$ can be chosen
by adjusting the density $\varrho$ of the medium and the spectral position of the probe light
frequency.

Apart from potential imaging applications the 5-level quantum interference scheme allows for 
devices operating in a wide range of positive and negative refractive indices
with simultaneously small losses.


\section{Impedance matching}
\label{ch:impedance}


When considering the applicability of optical devices reflection at boundaries 
between different media play an important role. Impedance matching at these
boundaries is often essential for the performance of the device.
For sub-wavelength resolution imaging the elimination of reflection losses
is particularly important, see \eq{eq:finetuning}.
In this section we thus derive conditions under which the boundary between non-chiral
and chiral, negative refracting media is non or little reflecting.

We consider a boundary between a non-chiral medium 1 ($z<0$) with $\varepsilon_1$, $\mu_1$
and medium 2 ($z>0$) which employs a cross-coupling ($\varepsilon_2$, $\mu_2$, $\xi_{EH}$, $\xi_{HE}$).
We assume again a wave propagating in $z$-direction in a medium corresponding to the tensor structure
(\ref{eq:TensorStructureXi}) such that we can restrict to an effectively scalar theory
for, e.g., left circular polarization. 

We decompose the $\ep$ field component in region 1 into an incoming $E_i$ and a reflected part $E_r$
\begin{equation}
\Eb_1(\rb)=(E_ie^{ik_1z}+E_re^{-ik_1z})\ep
\end{equation}
($k_1=|\kb_i|=|\kb_r|$). In medium 2 ($z>0$) only a transmitted wave $E_t$ exists due
to the boundary condition at infinity
\begin{equation}
\Eb_2(\rb)=E_te^{ik_2z}\ep.
\end{equation}
The connection of these modes at the interfaces and similar ones for the magnetic field $\Hb(\rb)$
is established by the boundary conditions $\nb\times(\Eb_2-\Eb_1)=0$
and $\nb\times(\Hb_2-\Hb_1)=0$. At $z=0$ we find
\begin{equation}
E_i+E_r=E_t\qquad\qquad H_i+H_r=H_t.
\label{eq:boundaryCond}
\end{equation}
Moreover an independent set of equations is found from Maxwell's equations in Fourier space
together with the material equations (\ref{eq:Pendrychiral}).
For medium 1 we get
\begin{equation}
\begin{split}
\kb_i\times & \ep E_i e^{ik_1z} +\kb_r\times\ep E_re^{-ik_1z}  = \\
& \frac{\omega}{c}\mu_1\left(H_ie^{ik_1z}+H_re^{-ik_1z}\right)\ep.
\end{split}
\label{eq:MaterialEquationMaterial1}
\end{equation}
Noting that $\ez\times{\bf e}_\pm=\pm i{\bf e}_\pm$ holds, this simplifies for $z=0$
to the scalar equation
\begin{equation}
ik_1(E_i -E_r) =\frac{\omega}{c}\mu_1(H_i+H_r)
\label{eq:Medium1}
\end{equation}
where $\kb_i=-\kb_r=k_1\ez$ has been applied. Similarly we obtain for medium 2
\begin{equation}
ik_2E_t=\frac{\omega}{c}(\xi_{HE}E_t+\mu_2 H_t).
\label{eq:Medium2}
\end{equation}
We eliminate the magnetic field amplitudes from (\ref{eq:boundaryCond}), (\ref{eq:Medium1}),
and (\ref{eq:Medium2}) and solve for the ratio of reflected and incoming electric field
amplitudes which yields
\begin{equation}
\frac{E_r}{E_i}=\dfrac{1-\sqrt{\dfrac{\mu_1}{\varepsilon_1}}\dfrac{n_2+i\xi_{HE}}{\mu_2}}{1+\sqrt{\dfrac{\mu_1}{\varepsilon_1}}\dfrac{n_2+i\xi_{HE}}{\mu_2}}.
\label{eq:Fresnel}
\end{equation}
Here the wave numbers $k_1$ and $k_2$ have been replaced by
$k_1=n_1\omega/c=\sqrt{\varepsilon_1\mu_1}\,\omega/c$ and $k_2=n_2\omega/c$.
Equation (\ref{eq:Fresnel}) is a generalization of the well-known Fresnel formulas for normal
incidence to a cross-coupled medium. Impedance matching is defined as the vanishing of the reflected
wave $E_r=0$, i.e., a complete transfer of the incoming field into medium 2:
\begin{equation}
\sqrt{\dfrac{\mu_1}{\varepsilon_1}}\dfrac{n_2+i\xi_{HE}}{\mu_2}=1.
\end{equation}
Using the explicit form of $n_2$ for the particular polarization mode (\ref{eq:refind})
we find the more convenient expression
\begin{equation}
\sqrt{\dfrac{\varepsilon_1}{\mu_1}}=\sqrt{\dfrac{\varepsilon_2}{\mu_2}}\left[\sqrt{1-\left(\dfrac{\xi_{EH}+\xi_{HE}}{2\sqrt{\varepsilon_2\mu_2}}\right)^2}+\dfrac{i}{2}\dfrac{\xi_{EH}+\xi_{HE}}{\sqrt{\varepsilon_2\mu_2}}\right]
\label{eq:impedanceMatch}
\end{equation}
%
\begin{figure}[t]
     \begin{center}
    \includegraphics[width=6.5cm]{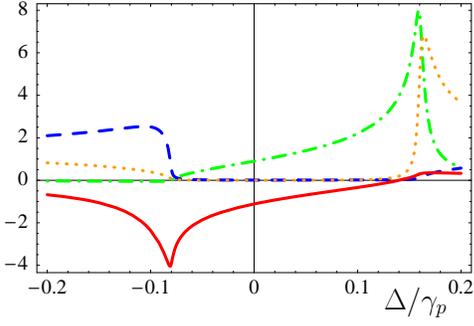}
    \caption{(color online) Real (solid line) and imaginary (dashed line) parts of the refractive index as well as the real (dash-dotted line) and imaginary (dotted line) parts  of the impedance $Z_2^{-1}$ for $\varrho=1,56\cdot 10^{17}$ cm$^{-3}$.}
       \label{fig:impedance}
     \end{center}
\end{figure}
which obviously simplifies to the well known result 
$\sqrt{\varepsilon_1/\mu_1}=\sqrt{\varepsilon_2/\mu_2}$ for the non-chiral case $\xi_{EH}=\xi_{HE}=0$.
The right hand side of (\ref{eq:impedanceMatch}) is the inverse impedance $Z_2^{-1}$
of the cross-coupled medium. For causality reasons the square root in $Z_2$ for passive media
has to be taken such that $\Re [Z_2]\geq0$ is obtained \cite{Smith2002}.
Fig.~\ref{fig:impedance} shows the real and imaginary parts of $Z_2^{-1}$ and of the index
of refraction. For the case $\varepsilon_1=\mu_1=1$ the impedances of the two media
are matched at the interface as soon as $Z_2^{-1}=1+i0$ holds. Applying the density
$\varrho=1.56\cdot 10^{17}$ cm$^{-3}$ we find $Z_2^{-1}=1.003+i0.0006$ at the spectral position 
$\Delta=1.17\cdot 10^{-2}\gamma_p$ while the index of refraction attains $n=-1.0003+i0.009$.
The corresponding figure of merit is about $\fom\approx 110$.


\section{Beyond 1D: Angle dependence}
\label{ch:2D}

\begin{figure}[t]
\begin{center}
\includegraphics[width=7cm]{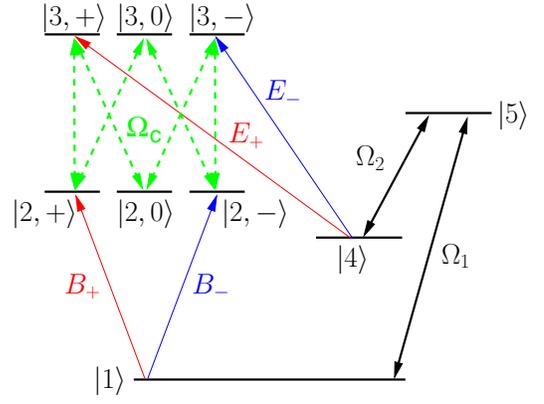}
   \caption{(color online) For a direction of incidence other than the direction of the coupling field vector $\Eb_c$ numerous additional angle dependent couplings occur.}
\label{fig:9level3D}
\end{center}
\end{figure}

In section \ref{ch:concepts} we specialized our discussion to an effectively scalar
theory by restricting to a particular direction of propagation and left circular polarization ($\ep$).
We now want to analyze the dependence of the refractive index on the propagation direction of the light, which
requires to take into account the tensor properties of all linear response coefficients.
To this end we consider a generalization of the 5-level scheme Fig.~\ref{fig:5levelpure}
that includes the full  Zeeman sublevel structure shown in 
Fig.~\ref{fig:9level3D}. The coupling field $\Eb_c$ is assumed to be linear polarized in the $z$-direction.
As the quantization axis we choose the propagation direction of the probe light.
In this scheme the requirements of section \ref{ch:concepts} are fulfilled.

Let us first consider the case when the probe light propagates along the $z$ axis.
Due to Clebsch-Gordon rules this solely leads to couplings $\Omega_c^{++}$ and $\Omega_c^{--}$
between the transitions $\ket{3,+} - \ket{2,+}$ and $\ket{3,-} - \ket{2,-}$, respectively.
As a result the two circular polarizations of a probe field traveling in $z$-direction
are eigenmodes and the scalar treatment from section \ref{ch:concepts} is valid.
We therefore get
\begin{equation}
n^{\pm}=\sqrt{\varepsilon\mu -\frac{\left(\xi_{EH}^\pm+\xi_{HE}^\pm\right)^2}{4}}\pm\frac{i}{2}\left(\xi_{EH}^\pm-\xi_{HE}^\pm\right)
\label{eq:refindpm}
\end{equation}
for the left [cf.~\eq{eq:refind})] and right circular polarizations.
The Wigner-Eckart theorem \cite{Cowan} implies that the electric dipole moments
of the $\ket{3,+} - \ket{4}$ and the $\ket{3,-} - \ket{4}$ transition coincide:
$d_{34}^+=d_{34}^-$. Similarly the matrix elements of the magnetic dipole transitions
are independent of the polarization state: $\mu_{21}^+=\mu_{21}^-$.
In contrast we find $d_{32}^{++}=-d_{32}^{--}$.
Thus the coupling Rabi frequencies of the left and right circular branches
$\Omega_c^{\pm\pm}=d_{32}^{\pm\pm} |\Eb_c|/\hbar$ have a relative sign
\begin{equation}
\Omega_c^{++}=-\Omega_c^{--}.
\end{equation}
From (\ref{eq:polEE}) - (\ref{eq:polBE}) together with the results obtained in section
\ref{ch:nonrad} we find the relations
\begin{equation}
\begin{split}
\varepsilon^+=\varepsilon^-\qquad & \qquad\mu^+=\mu^- \\
\xi_{EH}^+=-\xi_{EH}^-\qquad & \qquad\xi_{HE}^+=-\xi_{HE}^-
\end{split}
\label{eq:leftrightsigns}
\end{equation}
One recognizes that the refractive indices
of $\ep$- and ${\bf e}_-$-polarizations are identical
\begin{equation}
n^+=n^-.
\end{equation}
Hence the index of refraction of the scheme from Fig.~\ref{fig:9level3D} is independent
of the polarization state of probe light propagating in $z$-direction.
For this reason the electromagnetically induced cross-coupling in the scheme of Fig.~\ref{fig:9level3D}
does not correspond to a chiral medium for which the circular components should have different
refractive indices.

Next we allow for an angle between the propagation direction of the probe field and the
direction of the coupling field vector $\Eb_c$. In particular we employ a frame of reference
whose $z$-direction is fixed to the $\kb$-vector: $\kb\sim\ez$. The direction of the coupling field
which is fixed with regard to the laboratory frame is then given by polar angles $\theta$, $\phi$:
\begin{equation}
(E_{c,x},E_{c,y},E_{c,z})=|\Eb_c|(\sin\theta\cos\phi,\sin\theta\sin\phi,\cos\theta).
\label{eq:angleFields}
\end{equation}
As the atomic quantization axis is assumed to be given by the $\ez$-axis of the $\kb$-frame,
the probe field will encounter an unchanged
atomic level structure irrespective of the direction of propagation.
As indicated in Fig.~\ref{fig:9level3D} the $\ket{1} - \ket{4}$
transition is assumed to be a $J=0,M=0$ to $J=0,M=0$ transition and thus the dark state is
spherically symmetric and does not depend on the polar angles $\theta$ and $\phi$.

In this framework angle-dependent propagation is tak\-en into account by means
of angle dependent coupling Rabi frequencies
\begin{eqnarray}
\Omega_c^{++} &=&\Omega_{c,0}\cos\theta=-\Omega_c^{--},\\
\Omega_c^{+0} &=&\frac{\Omega_{c,0}}{\sqrt{2}}\sin\theta e^{-i\phi}=\Omega_c^{0-},\\
\Omega_c^{-0} &=&\frac{\Omega_{c,0}}{\sqrt{2}}\sin\theta e^{i\phi}=\Omega_c^{0+}
\end{eqnarray}
according to (\ref{eq:angleFields}). Here $\Omega_{c,0}=\left<3||\db||2\right>\cdot|\Eb_c|/(\sqrt{6}\hbar)$
is found from the Wigner-Eckart theorem where $\left<3||\db||2\right>$ denotes the reduced dipole
matrix element.

The angle-dependent cross coupling tensor for the electrically induced magnetization reads
\begin{equation}
\bar\alpha^{BE}=\alpha^{BE}\left(\begin{array}{ccc}
\cos\theta & 0 & \sin\theta\dfrac{e^{-i\phi}}{\sqrt{2}} \\[3mm]
0 & -\cos\theta & \sin\theta\dfrac{e^{+i\phi}}{\sqrt{2}} \\[3mm]
\sin\theta\dfrac{e^{+i\phi}}{\sqrt{2}} & \sin\theta\dfrac{e^{-i\phi}}{\sqrt{2}} & 0
\end{array}
\right)
\label{eq:PolarizationBEAngle}
\end{equation}
with $\alpha^{BE}$ given by (\ref{eq:polBE}).
Note that the tensor is expressed in the $\{+,-,z\}$-basis. For example the coefficient
$\alpha^{EB}_{+z}$ which describes the $\ep$-polarized electric field induced by a $\ez$-polarized
magnetic field (in the $\kb$-frame) is given by the upper right entry.
For $\bar\alpha^{EB}$ we find (\ref{eq:PolarizationBEAngle}) as well but with $\alpha^{BE}$
replaced by $\alpha^{EB}$ from equation (\ref{eq:polEB}).
On the other hand the electric polarizability is given by
\begin{equation}
\begin{split}
& \qquad\bar\alpha^{EE}= \alpha^{EE}\eins + \frac{\alpha^{EE}|\Omega_c|^2}{D_{42}D_{34}}\times \\[2mm]
& \times \left(\begin{array}{ccc}
\dfrac{\sin^2\theta}{8} & 
-\dfrac{\sin^2\theta\, e^{2i\phi}}{8} & 
-\dfrac{\sin\theta\cos\theta\, e^{i\phi}}{4\sqrt{2}} \\[3mm]
-\dfrac{\sin^2\theta\, e^{-2i\phi}}{8} & 
\dfrac{\sin^2\theta}{8} & 
\dfrac{\sin 2\theta\, e^{-i\phi}}{8\sqrt{2}} \\[3mm]
-\dfrac{\sin\theta\cos\theta\, e^{-i\phi}}{4\sqrt{2}} & 
\dfrac{\sin 2\theta\, e^{i\phi}}{8\sqrt{2}} & 
\dfrac{\cos^2\theta}{4}
\end{array}
\right)
\end{split}
\label{eq:PolarizationEEAngle}
\end{equation}
with $\alpha^{EE}$ determined by \eq{eq:polEE} and $D_{42}=(\gamma_{42}+i(\Delta_E-\delta_c))$,
$D_{34}=(\gamma_{34}+i\Delta_E)$. For the magnetic polarizability $\bar\alpha^{BB}$ the same
tensor structure applies. Again $\alpha^{EE}$ has to be replaced by (\ref{eq:polBB})
and $D_{42}D_{34}$ is substituted by
$D_{31}D_{21}=(\gamma_{31}+i(\Delta_B+\delta_c))(\gamma_{21}+i\Delta_B)$.

For incidence in the $z$-direction the tensors simplify significantly. The
cross-couplings reduce to a tensor proportional to $\ep\otimes{\bf e}_+^\ast-{\bf e}_-\otimes{\bf e}_-^\ast$
which identically corresponds to (\ref{eq:TensorStructureXi}) for $\xi^z=0$ and $\xi^+=-\xi^-$.
In the same limit ($\theta=0$) the electric and magnetic polarizabilities become diagonal.
In particular $\alpha^{EE}_{++}$ and $\alpha^{EE}_{--}$ are given by (\ref{eq:polEE})
and therefore potentially display EIT while the $\alpha^{EE}_{zz}$ entry simplifies to a
simple Lorentzian resonance structure.
In contrast the diagonal elements of $\bar\alpha^{BB}$ always display a Lorentzian resonance
with ($\alpha^{BB}_{++}$, $\alpha^{BB}_{--}$) and without ($\alpha^{BB}_{zz}$) coherent coupling.

From the angle dependent response tensors we find an angle dependent index of refraction.
The true index of refraction which takes the full form of (\ref{eq:PolarizationEEAngle}) into account,
i.e., the angle dependent correction to $\varepsilon$ and $\mu$,
gets very complicated. We here note that under the assumption of isotropic permittivity and
permeability we find the fairly simple result
\begin{equation}
\begin{split}
n^+=n^-=& \sqrt{\epsilon\mu-\xi_{EH}\xi_{HE}-\frac{(\xi_{EH}-\xi_{HE})^2\cos^2(\theta)}{4}}\\
& +\frac{i}{2}(\xi_{EH}-\xi_{HE})\cos(\theta).
\end{split}
\label{eq:nAngleDependentnochmal}
\end{equation}
independent of the polarization state. We conclude that even the idealized case
$\bar\varepsilon\sim\eins\sim\bar\mu$ does not give an isotropic index of refraction.
\begin{figure}[t]
\begin{center}
\includegraphics[width=6.5cm]{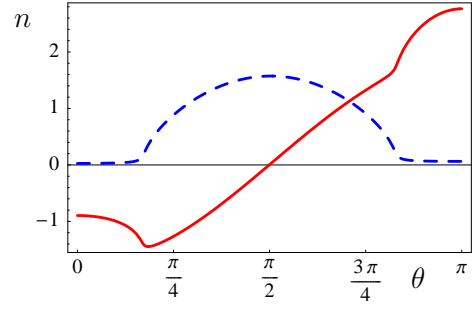}
     \caption{(color online) Real (solid line) and imaginary (dashed line) parts of the refractive index (\ref{eq:nAngleDependentnochmal}) as a function of $\theta$.}
\label{fig:aniso}
\end{center}
\end{figure}

In Fig.~\ref{fig:aniso} we show the index of refraction (\ref{eq:nAngleDependentnochmal})
as a function of the polar angle $\theta$. We use values of the response functions taken at
a spectral position $\Delta=-0,035\gamma_p$ for a density $\varrho=5\cdot 10^{16}\text{cm}^{-3}$.
We emphasize that the angle dependence in (\ref{eq:nAngleDependentnochmal}) results in an index
of refraction which varies over a broad spectrum of positive and negative values for different angles.


\section{Conclusion}
\label{ch:conclusion}

In conclusion we have shown that coherent magneto-electric cross-coupling
improves the prospects to obtain low-loss negative refraction in several ways.
The densities needed to get $\Re[n]<0$ are small enough to consider implementations
in, e.g., doped crystals. The presence of quantum interference effects
similar to electromagnetically induced transparency suppresses absorption and at the same time
enhances the magneto-electric cross-coupling. As a result our scheme allows for a tunable
low-loss negative refraction which can be impedance matched by means of external laser
fields.

\begin{acknowledgments}
M.F. and J.K. thank the Institute for Atomic, Molecular and
Optical Physics at the Harvard-Smithsonian Center for Astrophysics and
the Harvard Physics Department for their hospitality and support.
R.W. thanks D. Phillips for useful discussions.
J.K. acknowledges financial support by the Deutsche Forschungsgemeinschaft
through the GRK 792 ``Nichtlineare Optik und Ultrakurzzeitphysik''
and by the DFG grant FL 210/14.
S.Y. thanks the NSF for support.
\end{acknowledgments}


\renewcommand{\theequation}{\Alph{section}.\arabic{equation}}
\setcounter{equation}{0}
\setcounter{section}{0}
\renewcommand{\thesection}{Appendix \Alph{section}}

\section{Exact numerical solution of the Liouville equation}
\label{app:B}

To solve the Liouville equation to all orders in the 
probe field amplitudes $E$ and $B$ we first 
transform to a rotating frame and specialize to steady
state solutions. This gives a set of $25$ algebraic equations
which we cast into a matrix form by arranging the 5 diagonal $\rho_{11}\dots\rho_{55}$ and 20 off-diagonal density matrix elements $\tilde\rho_{21}\dots$ into a 25-dimensional vector
$\vec\rho$. We end up with an inhomogeneous matrix equation
\begin{equation}
{\mathcal M}\vec\rho=\vec a
\label{eqB:matrixeq}
\end{equation}
where the inhomogeneity is given by the 25-dimensional vector $\vec a=(1,0,0,\dots)$.
The matrix ${\mathcal M}$ contains all couplings, detunings and decay rates for
the system in question.
To solve equation (\ref{eqB:matrixeq}) for the sought density matrix elements
$\tilde\rho_{34}$ and $\tilde\rho_{21}$ we have to invert $\mathcal M$ which can only be done
numerically after specifying explicit numbers for all parameters.

In general we find that $\tilde\rho_{34}$ and $\tilde\rho_{21}$
are functions of both, the electric and the magnetic field amplitude
\begin{equation}
\tilde\rho_{34}=f(E,B) \qquad\qquad \tilde\rho_{21}=g(E,B).
\label{eqB:functions}
\end{equation}
We emphasize that the analytical form of the functions $f(E,B)$ and $g(E,B)$ is unknown.
As we want to compare with the result of linear response theory we have to
bring (\ref{eqB:functions}) in the form of \eq{eq:alphas}
\begin{equation}
d_{34}\tilde\rho_{34}=\alpha^{EE}(E,B)E+\alpha^{EB}(E,B)B
\end{equation}
\begin{equation}
\mu_{21}\tilde\rho_{21}=\alpha^{BE}(E,B)E+\alpha^{BB}(E,B)B.
\end{equation}
In contrast to the linear response theory here we deal with the exact solution of the 
Liouville equation and therefore the polarizabilities are still functions of 
the fields $E$ and $B$. At first glance the separation does not seem to be unique.
To determine $\alpha^{EE}(E,B)$, $\alpha^{EB}(E,B)$, $\alpha^{BE}(E,B)$, and $\alpha^{BB}(E,B)$
numerically we formally expand $f$ and $g$ in a power series in $E$ and $B$
\begin{equation}
f(E,B)=\sum_{n,m}f_{nm}E^nB^m
\end{equation}
\begin{equation}
g(E,B)=\sum_{n,m}g_{nm}E^nB^m.
\end{equation}
To separate the electric and magnetic properties uniquely we use the fact that physically
there must be an odd power of field amplitudes.
In fact all but one (fastly rotating) factors $e^{-i\omega_p t}$
must be compensated by factors $e^{i\omega_p t}$.
Otherwise the (untransformed) polarizabilities would not oscillate with the probe field
frequency $\omega_p$.
Since an odd power of field amplitudes can only be realized by an odd power in $E$ and an even
power in $B$ or vice versa an even power in $E$ and an odd power in $B$
we formally split into even and odd subseries
\begin{equation*}
f(E,B)=\sum_{n,m}f_{nm}^E|E|^{2n}|B|^{2m} E+\sum_{n,m}f_{nm}^B|E|^{2n}|B|^{2m} B,
\end{equation*}
\begin{equation*}
g(E,B)=\sum_{n,m}g_{nm}^E|E|^{2n}|B|^{2m} E+\sum_{n,m}g_{nm}^B|E|^{2n}|B|^{2m} B.
\end{equation*}
The polarizabilities are therefore given by the appropriate subseries, e.g.
\begin{equation*}
\alpha^{EE}(E,B)=d_{34}\sum_{n,m}f_{nm}^E|E|^{2n}|B|^{2m},
\end{equation*}
and similarly for $\alpha^{EB}(E,B)$, $\alpha^{BE}(E,B)$ and $\alpha^{BB}(E,B)$.
The utilization of symmetry properties then gives
\begin{equation}
\alpha^{EE}(E,B)=\frac{d_{34}}{2E}[f(E,B)+f(E,-B)]
\end{equation}
\begin{equation}
\alpha^{EB}(E,B)=\frac{d_{34}}{2B}[f(E,B)+f(-E,B)]
\end{equation}
\begin{equation}
\alpha^{BE}(E,B)=\frac{\mu_{21}}{2E}[g(E,B)+g(E,-B)]
\end{equation}
\begin{equation}
\alpha^{BB}(E,B)=\frac{\mu_{21}}{2B}[g(E,B)+g(-E,B)]
\end{equation}
which represents a unique numerical solution for the polarizability coefficients.


\def\etal{\textit{et al.~}}


\begin{thebibliography}{99}

\bibitem{Veselago68}
V. G. Veselago, Sov. Phys. Usp. {\bf 10}, 509 (1968).

\bibitem{Pendry99}
J. B. Pendry \etal, IEEE Trans. Micro. Theory Tech. {\bf 47}, 2075 (1999).

\bibitem{Smith00}
D. R. Smith, W. J. Padilla, D. C. Vier, S. C. Nemat-Nasser, S. Schultz, Phys. Rev. Lett. {\bf 84}, 4184 (2000).

\bibitem{Shelby01}
R. A. Shelby, D. R. Smith, S. Schultz, Science {\bf 292}, 77 (2001).

\bibitem{Shalaev07}
V. M. Shalaev, Nature Photonics {\bf 1}, 41 (2007).

\bibitem{Soukoulis07}
C. M. Soukoulis, S. Linden, M. Wegener, Science {\bf 315}, 47 (2007).

\bibitem{Yen04}
T. J. Yen \etal, Science {\bf 303}, 1494 (2004).

\bibitem{Linden04}
S. Linden \etal, Science {\bf 306}, 1351 (2004). 

\bibitem{Enkrich05}
C. Enkrich \etal, Phys. Rev. Lett. {\bf 95}, 203901 (2005).

\bibitem{Parimi04}
P. V. Parimi \etal, Phys. Rev. Lett. {\bf 92}, 127401 (2004).

\bibitem{Berrier04}
A. Berrier \etal, Phys. Rev. Lett. {\bf 93}, 073902 (2004).

\bibitem{Lu05}
Z. Lu \etal, Phys. Rev. Lett. {\bf 95}, 153901 (2005).

\bibitem{Yuan07}
H.-K. Yuan \etal, Optics Express {\bf 15}, 1076 (2007).

\bibitem{Shalaev05}
V. M. Shalaev \etal, Opt. Lett. {\bf 30}, 3356 (2005).

\bibitem{Klar06}
T. A. Klar, A. V. Kildishev, V. P. Drachev, V. M. Shalaev, 
IEEE J. Sel. Top. Qua. Electonics {\bf 12}, 1106 (2006).

\bibitem{Dolling07}
G. Dolling, M. Wegener, C. M. Soukoulis, S. Linden, Opt. Lett. {\bf 32}, 53 (2007).

\bibitem{Zhang05}
S. Zhang \etal, Phys. Rev. Lett. {\bf 95}, 137404 (2005).

\bibitem{Dolling07b}
G. Dolling, M. Wegener, C. M. Soukoulis, S. Linden, Opt. Express {\bf 15}, 11536 (2007).

\bibitem{Pendry00}
J. B. Pendry, Phys. Rev. Lett. {\bf 85}, 3966 (2000).

\bibitem{Leonhardt06}
U. Leonhardt, Science {\bf 312}, 1777 (2006).

\bibitem{Pendry06}
J. B. Pendry, D. Schurig, D. R. Smith, Science {\bf 312}, 1780 (2006).

\bibitem{Schurig06}
D. Schurig \etal, Science {\bf 314}, 977 (2006).

\bibitem{Smith03}
D. R. Smith \etal, Appl. Phys. Lett. {\bf 82} 1506 (2003).

\bibitem{Merlin}
R. Merlin, Appl. Phys. Lett. {\bf 84}, 1290 (2004).

\bibitem{DollingRekord}
G. Dolling, C. Enkrich, M. Wegener, C. M. Soukoulis, S. Linden, Opt. Lett. {\bf 31}, 1800 (2006).

\bibitem{Scully-chirality}
V. A. Sautenkov \etal, Phys. Rev. Lett. {\bf 94}, 233601 (2005).

\bibitem{Kaestel07a}
J. K\"astel, M. Fleischhauer, S. F. Yelin, R. L. Walsworth, Phys. Rev. Lett. {\bf 99}, 073602 (2007).

\bibitem{Fleischhauer05}
M. Fleischhauer, A. Imamoglu, and J. P. Marangos, Rev. Mod. Phys. {\bf 77}, 633 (2005).

\bibitem{Harris-Physics-Today-1997}
S. E. Harris, {\it Electromagnetically induced transparency}, Physics Today {\bf 50}, 36 (1997).

\bibitem{Kong1972}
J. A. Kong, Proc. IEEE {\bf 60}, 1036 (1972); J. A. Kong, J. Opt. Soc. Am. {\bf 64}, 1304 (1974).

\bibitem{ODell}
T. H. O'Dell, {\it The electrodynamics of magneto-electric media}, North-Holland Publishing Group (1970).

\bibitem{Pendry04}
J. B. Pendry, Science {\bf 306}, 1353 (2004).

\bibitem{Harris-NLO-EIT}
S. E. Harris, J. E. Field, and A. Imamoglu, Phys. Rev. Lett. {\bf 64}, 1107 (1990).

\bibitem{Stoicheff-Hakuta}
K. Hakuta, L. Marmet, and B. P. Stoicheff, Phys. Rev. Lett. {\bf 66}, 596 (1991).

\bibitem{Oktel04}
M. \"O. Oktel, and \"O. E. M\"ustecaplioglu, Phys. Rev. A {\bf 70}, 053806 (2004).

\bibitem{Thommen-PRL-2006}
Q. Thommen, and P. Mandel, Phys. Rev. Lett. {\bf 96}, 053601 (2006).

\bibitem{KastelComment}
Concerning \cite{Thommen-PRL-2006} see also J. K\"astel, M. Fleischhauer, Phys. Rev. Lett. {\bf 98}, 069301 (2007).

\bibitem{Cowan}
R. D. Cowan. {\it The theory of atomic structure and spectra}, 
University of California Press (1981).

\bibitem{Louisell}
W. H. Louisell, {\it Quantum statistical properties of radiation}, John Wiley \& Sons (1990).

\bibitem{Jackson}
J. D. Jackson, {\it Classical Electrodynamics},
Wiley (New York).

\bibitem{Cook}
D. M. Cook, {\it The Theory of the Electromagnetic Field},
Prentice-Hall (New Jersey).

\bibitem{Kaestel07}
J. K\"astel, M. Fleischhauer, and G. Juzeli\=unas, Phys. Rev. A {\bf 76}, 062509 (2007).

\bibitem{Smith2002}
D. R. Smith, S. Schultz, P. Marko\v{s}, C. M. Soukoulis, Phys. Rev. B {\bf 65}, 195104 (2002).


\end{thebibliography}
\end{document}